\documentclass[a4paper]{article}

\usepackage[english]{babel}
\usepackage[utf8x]{inputenc}
\usepackage[T1]{fontenc}

\usepackage[a4paper,top=3cm,bottom=2cm,left=3cm,right=3cm,marginparwidth=1.75cm]{geometry}

\usepackage{amsmath}
\usepackage{algpseudocode}
\usepackage{graphicx}
\usepackage{epsfig}
\usepackage{subcaption}
\usepackage[colorinlistoftodos]{todonotes}
\usepackage[colorlinks=true, allcolors=blue]{hyperref}

\title{CoHSI III:  Long proteins and implications for protein evolution}
\author{Les Hatton\footnote{Emeritus Professor, Kingston University, KT1 2EE, U.K., lesh@oakcomp.co.uk}, Gregory Warr\footnote{Emeritus Professor, Medical University of South Carolina, 96 Jonathan Lucas St, Charleston, SC 29425, USA, gwawarr@gmail.com}}

\begin{document}
\maketitle

\begin{abstract}
The length distribution of proteins measured in amino acids follows the CoHSI (Conservation of Hartley-Shannon Information) probability distribution.  In previous papers we have verified various predictions of this using the Uniprot database but here we explore a novel predicted relationship between the longest proteins and evolutionary time.  We demonstrate from both theory and experiment that the longest protein and the total number of proteins are intimately related by Information Theory and we give a simple formula for this.  We stress that no evolutionary explanation is necessary; it is an intrinsic property of a CoHSI system.  While the CoHSI distribution favors the appearance of proteins with fewer than 750 amino acids (characteristic of most functional proteins or their constituent domains) its intrinsic asymptotic power-law also favors the appearance of unusually long proteins; we predict that there are as yet undiscovered proteins longer than 45,000 amino acids. In so doing, we draw an analogy between the process of protein folding driven by favorable pathways (or funnels) through the energy landscape of protein conformations, and the preferential information pathways through which CoHSI exerts its constraints in discrete systems.

Finally, we show that CoHSI predicts the recent appearance in evolutionary time of the longest proteins, specifically in eukaryotes because of their richer unique alphabet of amino acids, and by merging with independent phylogenetic data, we confirm a predicted consistent relationship between the longest proteins and documented and potential undocumented mass extinctions.  
\end{abstract}

\section{Statement of computational reproducibility}
There is a growing awareness of the problem of computational irreproducibility in the software-consuming sciences, and so as with our previous papers in this area \cite{HatTSE14,HattonWarr2015,HattonWarr2017}, this paper is accompanied by a complete computational reproducibility suite including all software source code, data references and the various glue scripts necessary to reproduce each figure, table and statistical analysis and then regress local results against a gold standard embedded within the suite to help build confidence in the theory and results we are reporting.  This follows the methods broadly described by \cite{Ince2012} and exemplified in a tutorial and case study \cite{HattonWarr2016}.  These reproducibility suites are currently available at http://leshatton.org/ until a suitable public archive appears, to which they can be transferred.

\section{Introduction}
The diversity of life that has evolved on earth could not have done so without the successive emergence of novel proteins to carry out newly essential functions. It is axiomatic (albeit requiring some qualification) that the sequence of a protein determines its structure and thus its function (\cite{Anfinsen1973,Dill1042}). The function and evolution of proteins pose, on the surface, two major questions associated with the astronomical size of theoretical protein sequence space and, even for a given protein of average length, the vast number of possible conformational states it could explore before finding the "correct" (i.e. native) functional conformation (\cite{Levinthal1969,Dryden2008}. In recent decades these questions have been simplified by considerations that minimize e.g. the effective size of the amino acid alphabet and that have engaged polymer theory to consider protein folding in terms of energy landscapes (\cite{Ben-Naim2012,Dryden2008,GhoshDill2009}). However, the constraints imposed by information theory (through the Conservation of Hartley-Shannon Information or CoHSI) on protein sequence space, and its implications for protein evolution and function, have hitherto been unexplored.

We know a great deal about how variation arises in protein-encoding genes, how processes at the cellular and population level lead to the spread and fixation of novel genes in populations, and how speciation occurs. It is generally accepted that we have a good (if continually developing) understanding of the biological processes that underpin evolution. Thus, we are under no illusions that the idea we are introducing here, that the emergence of novelty in protein sequences is constrained by a conservation principle (CoHSI) arising from information theory, will be controversial. 

Information can be defined in different (but valid) ways, and information theory has been applied to gain insight into many aspects of biology (e.g. \cite{Frank2009,Vinga2014}) including evolutionary biology (e.g. \cite{Adami2012}). Often the starting point for these studies is to ask how the information present in a system  can be used to infer functional or evolutionary relationships, e.g. quoting from \cite{Adami2012} "Information stored in a biological organism's genome is used to generate the organism and to maintain and control it." Our approach to the implications of information theory for proteins is quite (indeed totally) different from this; it arises from an interest in the fundamental properties of discrete systems \textit{as a whole,} i.e. systems that consist of components each of which is made from smaller indivisible pieces (called tokens.)

Discrete systems are ubiquitous; they include e.g. the elements, software, texts, musical compositions, DNA and RNA and proteins. Our interest was therefore not to understand the role of information contained or indeed "used" in any particular discrete system, but rather to understand how fundamental information theory might constrain the properties of \textit{all} discrete systems. Thus, to be applicable to proteins as well as software, texts and musical compositions etc. the theory has to be token agnostic, i.e. eschewing any meaning or function associated with the tokens; we therefore used information in the sense of Hartley-Shannon Information, i.e. a change of sign without  reference to any meaning. Indeed Hartley specifically cautioned against associating \textit{any meaning} to the tokens \cite{Hartley1928}. We emphasize that in both Hartley-Shannon Information and the Statistical Mechanical framework in which it is embedded in our theory that the \textit{type} of token has no relevance. This is quite counter-intuitive when we are used to mechanisms e.g. looking at proteins from the perspective of protein chemistry and biochemical function, but CoHSI has a completely different perspective, considering any discrete system as an ergodic ensemble of components unrestrained with respect to space or time and characterised only by equi-probable microstates. 

Embedding Hartley-Shannon information \cite{Hartley1928,Shannon1948} in a Statistical Mechanical framework \cite{HattonWarr2017,HattonWarr2018a} showed that the length distribution of components in a discrete system is overwhelmingly likely to obey a differential equation (the CoHSI equation) which implicitly defines the probability distribution function. Solving this equation \cite{HattonWarr2018a} gave a canonical distribution of component sizes that predicted the length distribution of components in any discrete system of sufficient size and complexity. When this prediction was examined experimentally, the length distribution of component sizes in such disparate systems as the total collection of known proteins, texts and the functions in computer programs (written in any programming language) were shown to conform accurately to the CoHSI distribution \cite{HattonWarr2018a}.  

Since proteins are a classical discrete system, and intimately involved with the evolution of life, it is worthwhile to ask here what constraints might be imposed by CoHSI on the evolution of protein novelty. We have already shown that the canonical CoHSI distribution is observable in proteins, as predicted, at all scales \cite{HattonWarr2017}.  As such, it constrains the distribution of protein lengths at taxonomic levels ranging from individual species through the domains of life and to the entirety of protein sequences documented in the databases.  We have also shown that the average length of proteins is highly constrained, as predicted by CoHSI \cite{HattonWarr2018b}.  Here we address a no less interesting property of proteins: an important prediction of the length distribution constrained by the CoHSI equation is that very long components are \textbf{inevitable}. Thus we can ask the question, is the longest protein in an aggregation \textit{directly predicted} by the total number of proteins in the aggregation?  If it is, what impact have the constraints imposed by CoHSI had on the emergence of novel protein sequences as life has evolved on earth?

\section{The CoHSI distribution and taxonomic level: scale independence }
From \cite{HatTSE14,HattonWarr2015,HattonWarr2017}  the fundamental CoHSI equation (1) was determined, the solution of which gives the canonical distribution of component lengths shown in Fig 1. 

\begin{equation}
\log t_{i} = -\alpha -\beta ( \frac{d}{dt_{i}} \log N(t_{i}, a_{i}; a_{i} ) ),    \label{eq:minifst}
\end{equation} 

where $t_{i}$ is the distribution of lengths and $N(t_{i}, a_{i}; a_{i})$ is the number of different ways in which the $t_{i}$ tokens in the $i^{th}$ component chosen from a unique alphabet of $a_{i}$ tokens can be arranged where order is important. The two undetermined Lagrange multipliers are $\alpha$ and $\beta$. For the proteins, the tokens are the amino acids comprising both the 22 directly encoded in the genome, as well as those which have been subjected to post-translational modification (PTM).

\begin{figure}[ht!]
\centering
\includegraphics[width=0.5\textwidth]{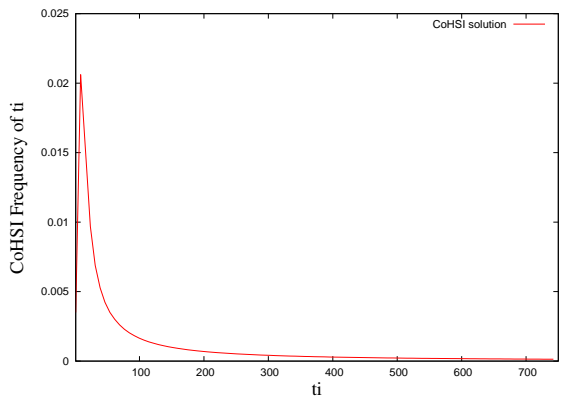}
\caption{\label{fig:mdata}Illustrating a typical solution of the CoHSI equation described in \cite{HattonWarr2017}. Both a sharp unimodal peak and power-law tail can be seen clearly}
\end{figure}

\begin{figure}[ht!]
\centering
\includegraphics[width=0.5\textwidth]{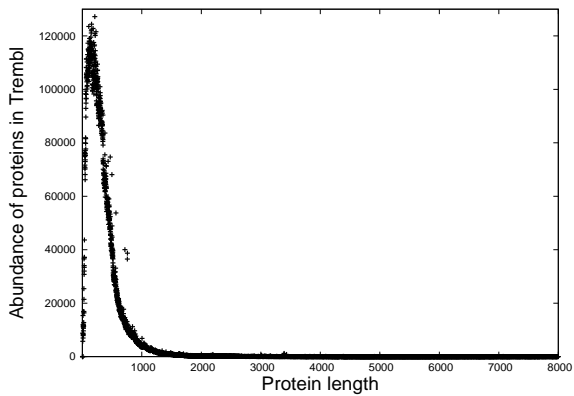}
\caption{\label{fig:tdata}Illustrating the length distribution of the proteins measured in amino acids in TrEMBL version 15-07.}
\end{figure}

The canonical distribution shown in Fig \ref{fig:mdata} can then be compared with the distribution (Fig \ref{fig:tdata}) of protein sizes (expressed as their length in amino acids) observed in a large aggregation of protein sequences (TrEMBL v15-07, https://uniprot.org/); the predicted sharp unimodal peak at small $t_{i}$ and  precise power-law tail of higher lengths of Fig \ref{fig:mdata} can be seen clearly in Fig \ref{fig:tdata} (see \cite{HattonWarr2017} for a detailed analysis of this).

\begin{figure*}[t!]
    \captionsetup[subfigure]{labelformat=empty}
    \centering
    \begin{subfigure}[t]{0.5\textwidth}
        \centering
        \caption{A}
        \includegraphics[width=6cm]{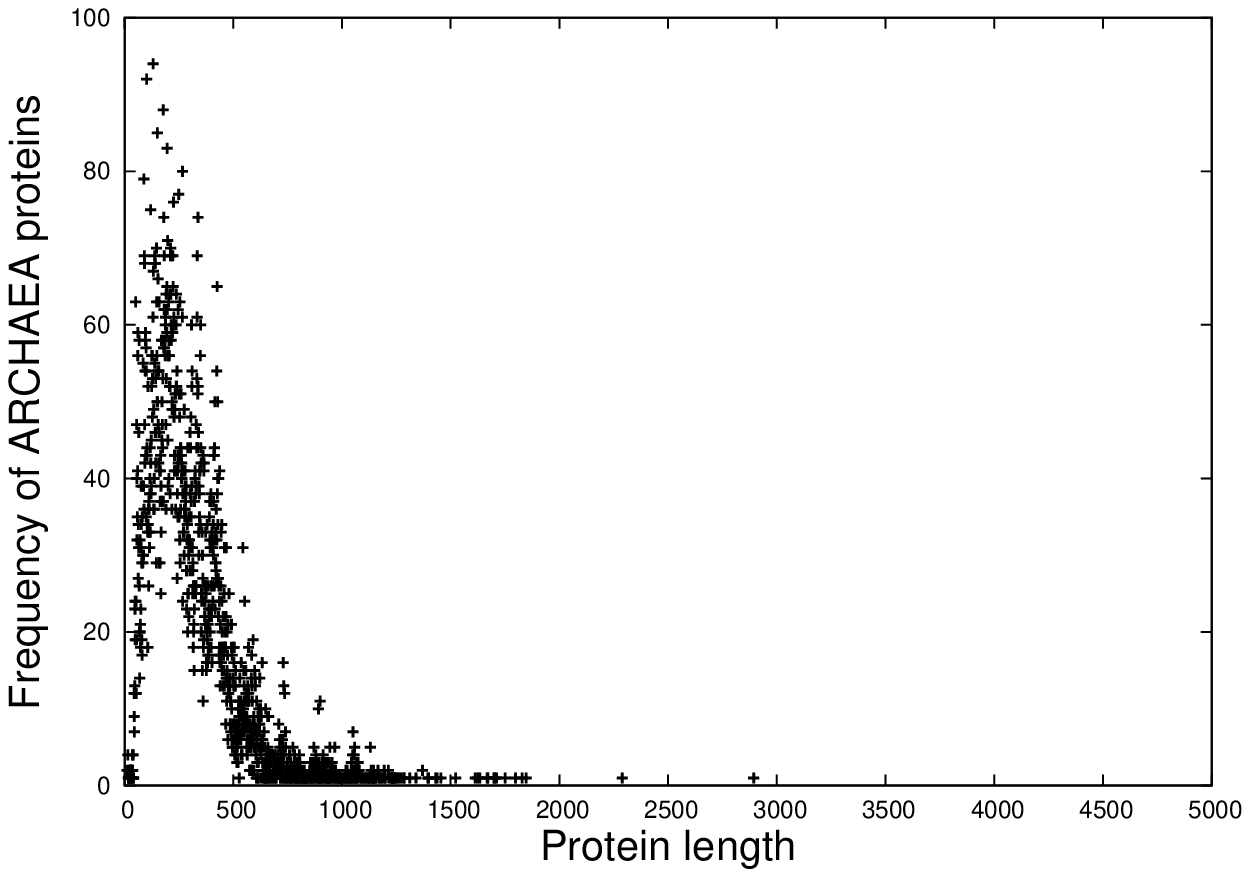}
        \label{fig:archaea_eps}
    \end{subfigure}%
    ~
    \begin{subfigure}[t]{0.5\textwidth}
        \centering
        \caption{B}
        \includegraphics[width=6cm]{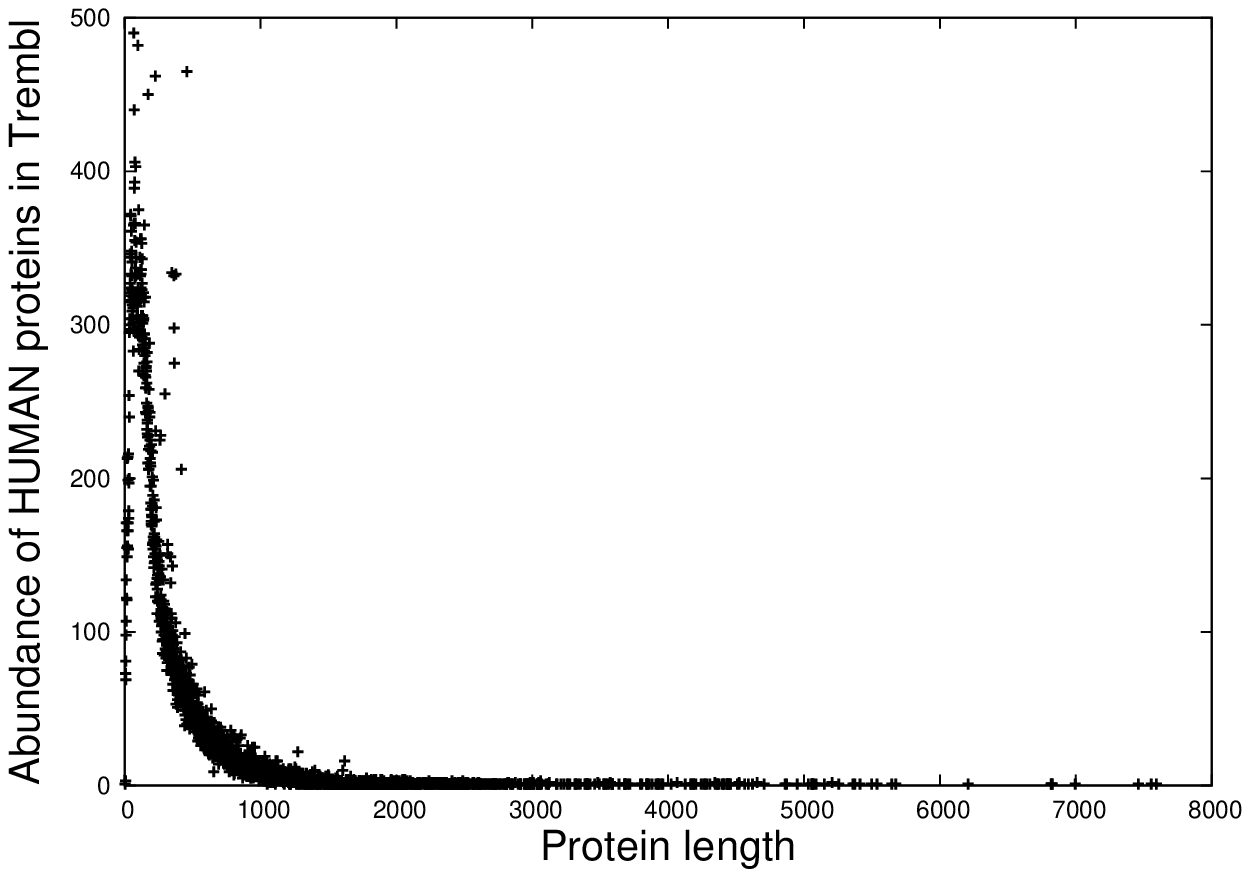}
        \label{fig:human_eps}
    \end{subfigure}%

    \begin{subfigure}[t]{0.5\textwidth}
        \centering
        \caption{C}
        \includegraphics[width=6cm]{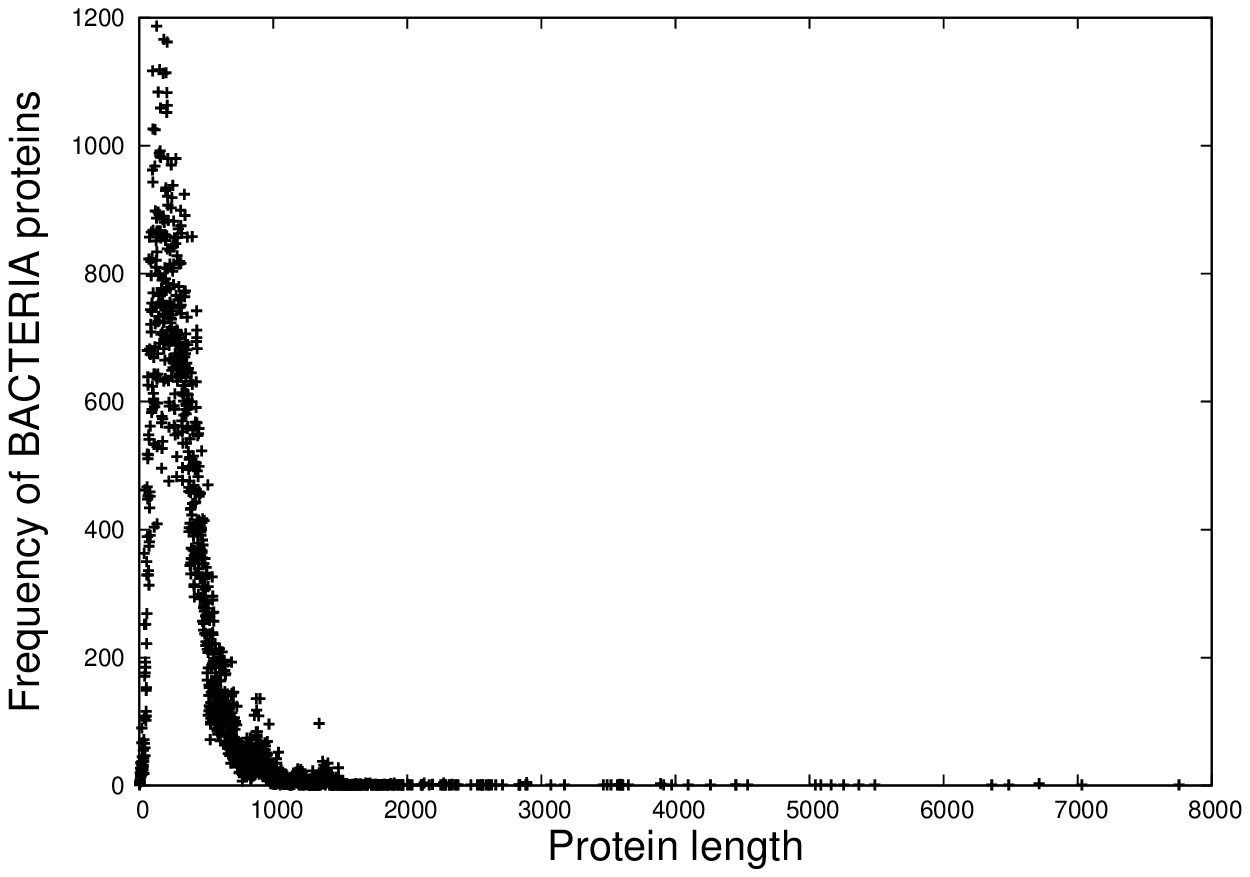}
        \label{fig:bacteria_eps}
    \end{subfigure}%
    ~
    \begin{subfigure}[t]{0.5\textwidth}
        \centering
        \caption{D}
        \includegraphics[width=6cm]{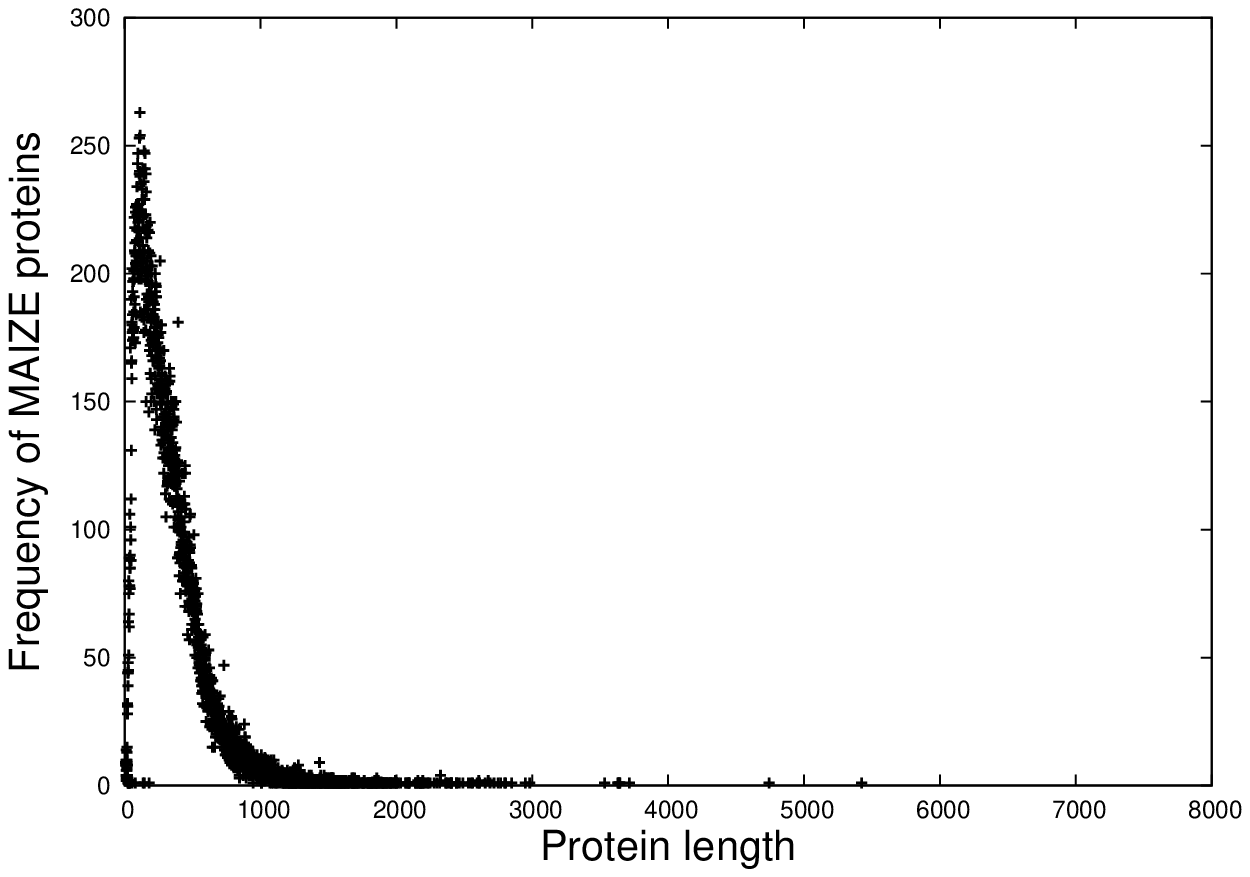}
        \label{fig:maize_eps}
    \end{subfigure}%

    \begin{subfigure}[t]{0.5\textwidth}
        \centering
        \caption{E}
        \includegraphics[width=6cm]{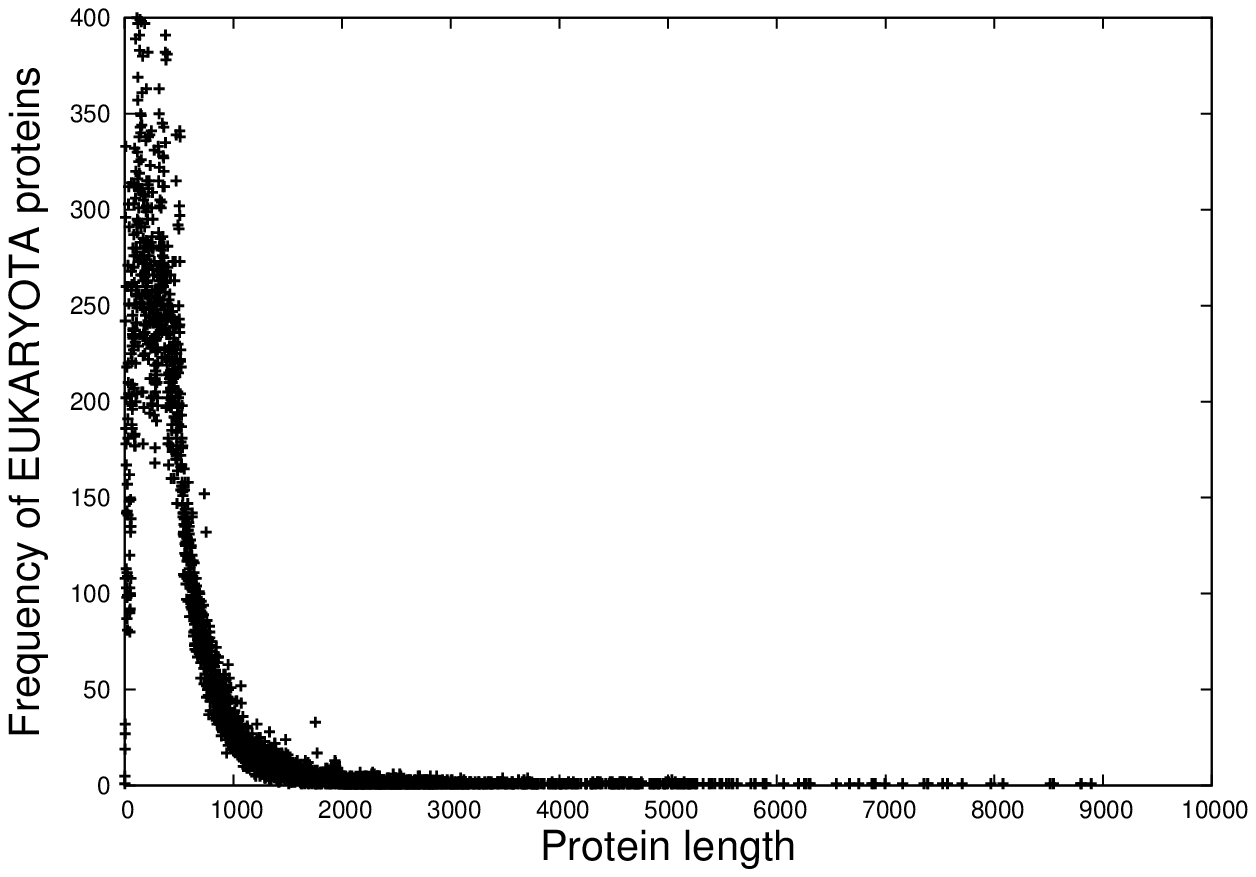}
        \label{fig:eukaryota_eps}
    \end{subfigure}%
    ~
    \begin{subfigure}[t]{0.5\textwidth}
        \centering
        \caption{F}
        \includegraphics[width=6cm]{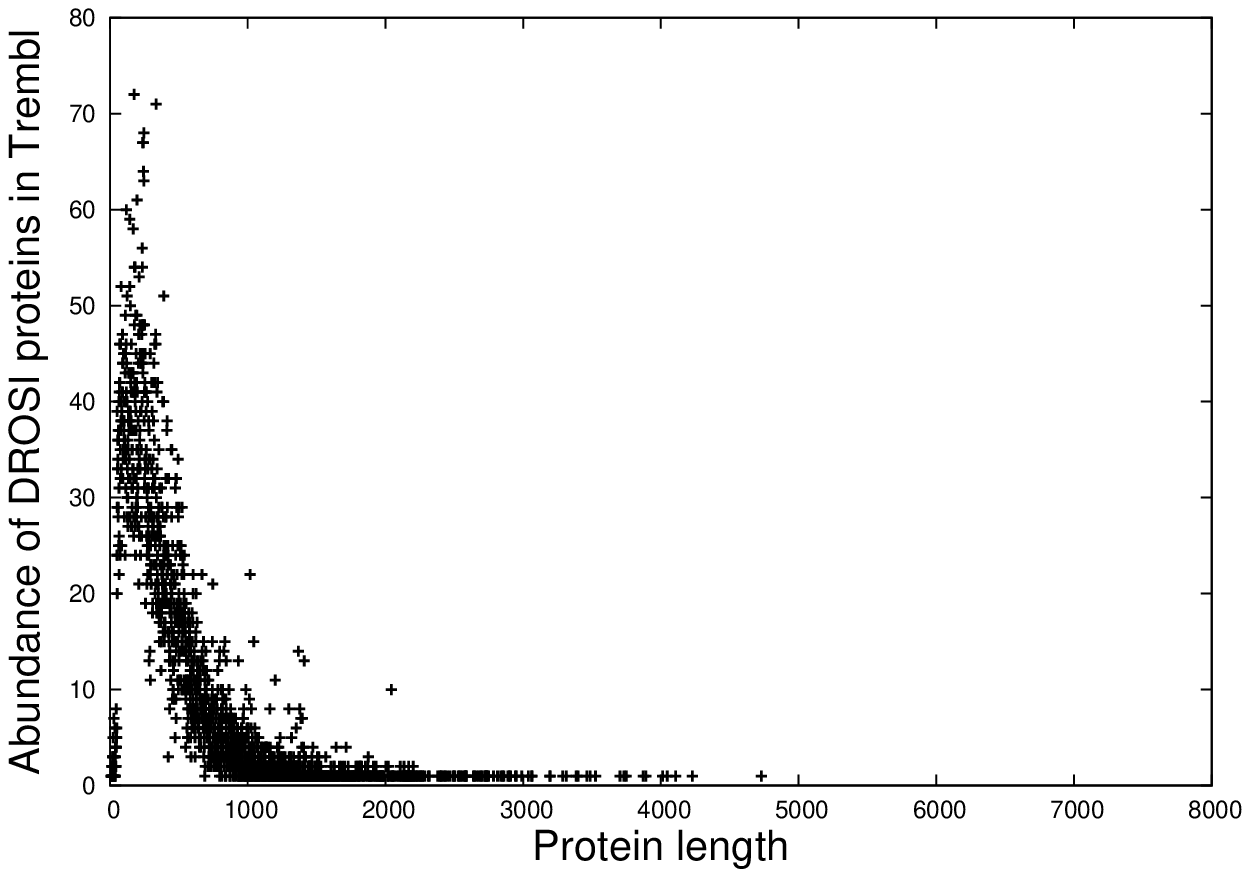}
        \label{fig:drosi_eps}
    \end{subfigure}%
    \caption*{Different aggregation sizes, all illustrating the CoHSI length distribution: A) Archaeal proteins; B) Human proteins; C) Bacterial proteins; D) Maize proteins; E) Eukaryotic proteins; and F) Fruit-fly proteins.}
\end{figure*}

We note that the canonical distribution of protein sizes can also be seen at scales ranging from the domains of life (archaea Fig \ref{fig:archaea_eps}, bacteria Fig \ref{fig:bacteria_eps}, eukaryota Fig \ref{fig:eukaryota_eps}) down to the level of species (human Fig \ref{fig:human_eps}, maize Fig \ref{fig:maize_eps}, and fruit fly Fig \ref{fig:drosi_eps}).

The presence of the power-law means that the distribution of protein sizes is long-tailed, and along with the sharp unimodal peak at small $t_{i}$, leads to a distribution which is palpably right-skewed.  This significantly complicates the meaning of the word \textit{average} as we discussed in detail in \cite{HattonWarr2018b}. By way of example and to show that common measures of the average such as mean, median and mode can differ substantially, if we compute each of these for the length distributions of the domains of life  in TrEMBL release 15-07 of Figure \ref{fig:tdata}, we get the values shown in Table \ref{tab:average}. The values for the mean lengths of proteins agree closely with those calculated by Kozlowski \cite{doi:10.1093/nar/gkw978} for example.  The effects of the skew can be clearly seen in the significant difference between mean and median.

\begin{table}[h!]
\centering
\caption{Measures of average length of proteins in the domains of life}
\begin{tabular}{@{\vrule height 10.5pt depth4pt  width0pt}cccc}
\hline
Domain of Life & Mean & Median & Mode \\
\hline
Archaea & 287 & 246 & 130 \\
Bacteria & 312 & 272 & 156 \\
Eukaryota & 435 & 350 & 379 \\
\hline
\end{tabular}
\label{tab:average}
\end{table}

The two undetermined Lagrange parameters $\alpha,\beta$ play an interesting role. From \cite{HatTSE14,HattonWarr2015,HattonWarr2017}, $\beta$ was shown to be intimately related to the range of unique token alphabet sizes encountered in a discrete system.  It is in fact the asymptotic slope of the power-law \textit{in the pdf of the unique alphabets,} $p_{i} \sim a_{i}^{-\beta}$.  For systems with bigger unique alphabets than the known collection of proteins, such as software systems where the tokens are programming language tokens, the $a_{i}$ can be in the hundreds or even thousands.  For such systems, the power-law is flatter corresponding to a smaller $\beta$, \cite{HatTSE14,HattonWarr2017}. For the proteins the alphabet ranges are, at least currently, much smaller primarily because of technological limitations in the measurement of the degree to which post-translational modification enhances the basic 22-letter amino acid alphabet directly coded from DNA.  For example, the largest unique alphabet of any protein in the SwissProt subset of TrEMBL release v.13-11 using the results of the Selene project \cite{Selene2013} was 33 \cite{HattonWarr2015}.  We expect the largest unique alphabet of any protein to increase markedly in size in the years to come, as the thousands of PTMs both known or predicted are better identified and annotated \cite{ApweilerHermjakobSharon1999,ZafarNasirBokhari2011}. This we expect to happen rather slowly as PTM  identification is time-consuming.  \textbf{We therefore expect $\beta$ to vary slowly  as the size of the collection of known proteins increases.}

Now, as we showed in \cite{HattonWarr2017,HattonWarr2018a}, the \textit{length distribution} of proteins measured in amino acids also asymptotes to a power-law, but whose slope $\beta'$ say, is related in a complex way to \textit{both} $\alpha$ and $\beta$.  However, this too is slowly varying so \textit{we expect that as the known total collection of proteins increases in size, the power-law in the tail of the length distribution will retain a nearly constant slope $\beta'$}.  In other words, the shape of the canonical CoHSI length distribution is essentially scale-invariant.  This is predicted from the nature of the CoHSI equation \cite{HattonWarr2017} and it is possible to test this prediction.

\begin{figure}[ht!]
\centering
\includegraphics[width=0.5\textwidth]{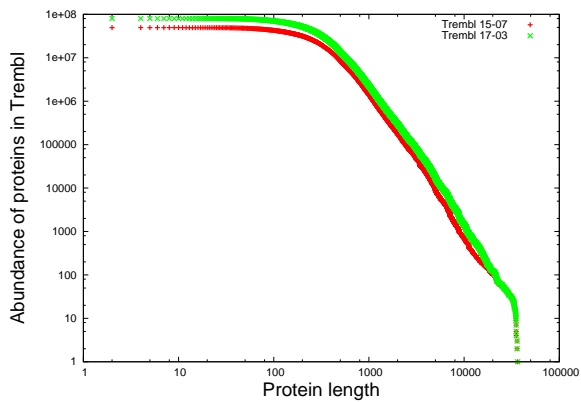}
\caption{\label{fig:tremblversions} The complementary cumulative distribution function (ccdf) of two versions of the full TrEMBL protein database separated by 20 months.}
\end{figure}

To see this in real data, Figure \ref{fig:tremblversions} shows the length distribution of two versions of the full TrEMBL protein database, versions 15-07 and 17-03 as a complementary cumulative distribution function (ccdf), in which the power-law has slope $-\beta'+1$, \cite{Newman2006}.  In the 20 months separating these two releases, the total number of proteins appearing in the database has increased by 60\% from around $5 \times 10^{7}$ to $8 \times 10^{7}$.  \textit{As predicted, the power-law slopes are both self-similar and emphatic.}  We will consider this as an equilibrium state for the system of proteins as we discussed in \cite{HattonWarr2018b}.

\begin{figure}[ht!]
\centering
\includegraphics[width=0.5\textwidth]{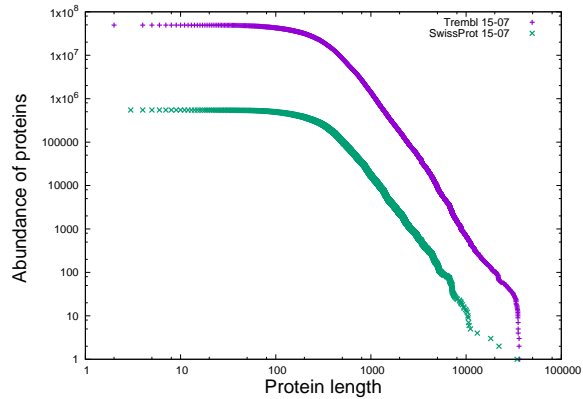}
\caption{\label{fig:swissprottrembl}Illustrating the complementary cumulative distribution function (ccdf) for the curated SwissProt subset and the full TrEMBL dataset in version 15-07, \cite{HattonWarr2017}.}
\end{figure}

This scale-independence is also clearly visible in subsets of the same dataset.  Figure \ref{fig:swissprottrembl} shows the full TrEMBL dataset, version 15-07, and the curated SwissProt subset of this version.  Even though the difference in size of the datasets is two orders of magnitude, the self-similarity is obvious as predicted.

Exactly the same phenomenon is observed in collections of software \cite{HatTSE14}.  Along with the highly conserved average component length  \cite{HattonWarr2018a}, this self-similarity is a natural property of all CoHSI systems and has a particularly interesting implication - the prediction that there is a relationship between the total number of components in an aggregation (the y-axis of Fig. \ref{fig:tremblversions}) and the maximum size of component within that aggregation (the x-axis of Fig. \ref{fig:tremblversions}). In the following section we test this prediction for proteins.

\section{A total size v. maximum length relationship}
Here we use a simple geometric argument based on self-similarity to derive a relationship between the total number of proteins in a collection and the maximum length of protein within that collection.

\begin{figure}[ht!]
\centering
\includegraphics[width=0.5\textwidth]{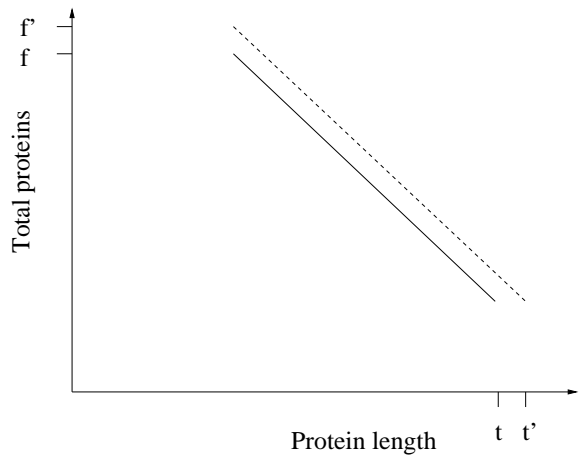}
\caption{\label{fig:geometry}A simple geometric model of the self-similarity of the protein length distribution represented as a ccdf.}
\end{figure}

Consider Figure \ref{fig:geometry}.  This is a ccdf schematic with logarithmic axes to base 10 so we can derive a simple geometric relationship by imagining a system growing in size from $f$ to $f'$ in the total number of proteins and a matching self-similar growth from $t$ to $t'$ in the maximum length of a protein in that population. We then get,

\begin{equation}
\frac{\log_{10} f - \log_{10} f'}{\log_{10} t' - \log_{10} t} = - \beta' + 1
\end{equation}

As mentioned earlier, we have used the fact that the power-slope on a cumulative distribution function is one greater than its slope on a pdf \cite{Newman2006}, which we have already taken to be $\beta'$.  Re-arranging, we derive the relationship

\begin{equation}
t' = t. \big [ (\frac{f'}{f})^{1/(\beta'-1)} \big ]
\label{eq:geometry}
\end{equation}

We can quantify this by inspecting Figure \ref{fig:tremblversions}.  The graphs are noisy for the largest protein itself but we can get a more robust idea by identifying the protein length \textit{after which} the longest 1000 proteins appear when arranged in increasing order.  For TrEMBL release 15-07, this length is 8,886 amino acids and for release 17-03 it is 10,787 amino acids.   This corresponds to a relative increase in length of $t'/t = 10787/8886 = 1.21$.

Moreover, there are $f = 49,01,998$ proteins in release 15-07 and $f' = 80,204,459$ proteins in release 17-03.  If we substitute these numbers into (\ref{eq:geometry}) for a value of the slope of around $\beta' = 4.13$ \cite{HattonWarr2017}, we get an estimate of $t'/t \approx 1.17$ which is satisfactorily close.

\textbf{We close this section by asserting therefore that the longest protein in a collection of proteins is constrained solely by the total number of proteins in that collection by (\ref{eq:geometry}).  No evolutionary mechanism is necessary, it is simply a general property of the CoHSI equation, its natural and precise asymptote to a power-law in length, and the relative constancy of the slope.}

\section{Possible constraints on the longest proteins}
In order to explore the implications of the assertion that the longest proteins of a collection will grow in a predictable way with the total number of proteins in that collection, we will first attempt to identify if current theory places any physico-chemical or combinatorial constraints on such growth.

\subsection{What physico-chemical bounds might constrain the longest protein size?}
An examination of theoretical work in protein folding does not appear to place any obvious bounds although much of the work has focused on globular proteins.  At the heart of this discussion is the Levinthal paradox \cite{Levinthal1969}, which states that an exhaustive search of all folding possibilities would lead to unreasonably long folding times for any reasonably long protein.  In a fascinating discussion Dryden et al \cite{Dryden2008} amongst other things address this question of how proteins can find their native folded conformation given the large (in principle astronomically vast) number of possible intermediate folding states. Using the concept of both unique amino acid alphabets and insensitivity to the actual amino acids used (but without reference to information content), they found that the effective number of possibilities could be greatly reduced. They made several assumptions, the most important of which is that for proteins to fold into a native state, the physico-chemical properties of the amino acid side chains can be reduced effectively to two classes, hydrophilic and hydrophobic. Such a reduced functional alphabet of amino acids (in our nomenclature $a_{i} = 2$) is based on the assumptions discussed by Dryden et al. (\cite{Dryden2008} that protein folding is driven essentially by hydrophobic effects  and that protein functionality is not necessarily constrained to unique sequences supported by the framework of the native protein.

Numerous other mechanisms have been proposed to resolve the Levinthal paradox as discussed in detail by \cite{Finkelstein2013}.  Moreover the existence of partitioning into semi-autonomous protein domains of 25-500 amino acids in length also ameliorates this paradox.  \cite{Finkelstein2013} go on to give a phenomenological formula for dependence of the folding rate on the size, shape and stability of protein folds.

Studies such as those by \cite{LanePande2013} show that for domains up to about 300 amino acids in length various rate-folding laws such as power-law and exponential fit experimental data well with corresponding folding times between around $10^{-8}s.$ and $10^{2}s$.  Using a theoretical argument \cite{GhoshDill2009}, based on thermodynamics, derive a relationship between protein stability and various physico-chemical properties such as temperature, denaturant, pH and salt.  Importantly, this relationship is linear in protein chain-length measured in amino acids rather than being non-linear.

Together these studies do not appear to throw up any insurmountable barriers to the gradually increasing length of the longest proteins that we predict based on the total number of proteins.

We will consider later in this discussion a possible link between the longest proteins in a given species with the point in evolutionary time when the species emerged,  invoking a bridging medium of the total number of proteins. Thus we need to explore the issue of the total number of proteins that have existed through the history of life on earth.

\subsection{What bounds might exist theoretically on the total number of proteins?}
The total possible population of proteins must by definition have grown from zero in early time, and despite its (arguable) monotonic growth as an ergodic system, acting on all proteins considered over time and space, the number of proteins present at any given time is a moot point given the history of successive mass extinction events \cite{Benton2005,Rothman2017}.  Clearly a precise knowledge of the number and diversity of proteins present throughout evolutionary history is without our power to obtain, and thus plausible assumptions need to be explored.

Dryden et al. \cite{Dryden2008} made assumptions regarding the upper and lower bounds for the number of novel proteins generated during 4Gyr of evolution. This lower bound is $4\times10^{21}$ proteins, and the upper bound is $4\times10^{43}$ proteins.  They claim that "most of the sequence space may have been explored" during evolution.  This then suggests that after an early phase when the sequence space was being explored by life, that the total number of proteins rose rapidly from zero to some value at which it remained stable. This is questionable. The authors' argument that the physico-chemical complexity of amino acid structures can be reduced to a unique alphabet as low as 2 or 3 is certainly challenging, but we also note that their argument is restricted to proteins of 100 amino acids or less in length; they consider that "The exploration of longer chains of 100 amino acids with only two types of residue is obviously much less complete but it is not a negligible fraction of the total".  This appears to be an overstatement.  The median length of proteins (Table 1) currently ranges from 246 amino acids for the archaea to 350 amino acids for the eukaryota \cite{HattonWarr2018b} implying by definition that half are longer. Even with the most parsimonious assumption of just $a_{i} = 2$ types of amino acid, a median length archaean protein offers $2^{246} = 10^{74}$ possible ways of arranging its amino acids and a median length eukaryotic protein offers $2^{350}$ = $2.29\times10^{105}$. 
 
As we will confirm when we consider the amount of protein sequence space which has been documented so far, this suggests that only a tiny fraction of this space could have been explored in evolutionary history.  It can be argued of course that we have no sequences for all the lifeforms that have disappeared but we must also consider that half of all sequences are longer than the median sequence and some of these, because of the slowly decaying power-law tail of the CoHSI distribution, are 1 to 2 orders of magnitude longer.

Let us examine more closely how CoHSI allows us a different perspective into the theoretical extent of relevant sequence space.  The CoHSI distribution has two defining features of relevance for the evolution of novelty in proteins, Fig \ref{fig:tdata}. On the one hand, the majority of proteins are smaller than 750 amino acids, whilst on the other, very long proteins (1 or 2 orders of magnitude larger than the median) are overwhelmingly likely to be present. We can recall \cite{HattonWarr2017} that CoHSI begins with a basic Statistical Mechanics formulation in which the total number of possible arrangements of $T$ tokens distinguishable by their order amongst $M$ components with $t_{i}$ tokens in the $i^{th}$ component is

\begin{equation}
\Omega(T,M) = \frac{T!}{\prod_{i=1}^{M} (t_{i}!)}		\label{eq:distinguishable}
\end{equation}

This is the total ergodic space of Statistical Mechanics. To render it a little less overwhelming, (even for modest systems $T!$ is astronomically large compared with the denominator), it is conventional to work with the logarithm of $\Omega(T,M)$ and employ Stirling's theorem yielding

\begin{equation}
\log \Omega(T,M) \sim T \log T - {\sum_{i=1}^{M} (t_{i} \log t_{i})}		\label{eq:logdistinguishable}
\end{equation}

By applying constraints to this, we define subsets that are the most likely pathways (or highways) through the information landscape of ergodic space, thus whittling down the number of possible configurations, i.e. the size of sequence space, in which systems can appear.  In the case of CoHSI, we constrain the possibilities by considering all those possible systems in time and space which have the same $T$ and a fixed $I$ where $I$ is the total Hartley-Shannon information split such that the information content of the $i^{th}$ component is $I_{i}$.  In this case, $I_{i}$ is simply the $N()$ function of equation (\ref{eq:minifst})

\begin{equation}
I_{i} = N(t_{i}, a_{i}; a_{i} )		\label{eq:hsi}
\end{equation}

The motivation for this is that the Statistical Mechanical methodology will then reveal the most likely length distribution which any system of given $T,I$ will have, exactly analogous to the Maxwell-Boltzmann distribution of classical kinetic theory when energy and size are the constraints \cite{GlazerWark2001}.

In the language of Statistical Mechanics then the length distributions governing these information highways through ergodic space are given by maximising

\begin{multline}
\log \Omega(T,M) = T \log T - T - \sum_{i=1}^{M} \lbrace t_{i} \log (t_{i}) - t_{i} \rbrace \\
+ \alpha \lbrace T - \sum_{i=1}^{M}t_{i} \rbrace + \beta \lbrace I -  \sum_{i=1}^{M} I_{i} \rbrace    \label{eq:mini}
\end{multline}

The solution of this process is of course equivalent to solving the differential equation (\ref{eq:minifst}) giving the range of distributions we explored in \cite{HattonWarr2018a}.  These distributions all asymptote to power-laws for components large compared with their unique alphabet of tokens, perhaps inviting the use of \textit{information superhighways} through sequence space.

\subsubsection{Information landscapes - an analogy with protein folding?}
The concept of an information landscape with preferential pathways  (highways or superhighways) that constrain the properties of discrete systems has an analogy with the energy landscapes that are associated with protein folding. \cite{Finkelstein2013} review numerous mechanisms which attempt to provide an answer to the question posed by Levinthal \cite{Levinthal1969}, of how proteins of even modest length can reach their final stable, correct (i.e. native) conformation given the astronomically large number of folding intermediates that are possible. This challenge is reduced to manageable proportions through the understanding that the process of folding offers an energy landscape in which energetically preferential (lower energy) states exist and by means of which pathways through this landscape (characterized by \cite{DillChan1997} as "funnels") channel the protein towards its native conformation as it folds.

We note however that strictly speaking, CoHSI and the methodology of Statistical Mechanics does not remove possible configurations - \textit{it is not a mechanism.}  All it does is recognize those configurations which because they have many more corresponding equi-probable microstates, are most likely to be seen in a given system.  As we have noted before, it is not a strait-jacket.  However, these configurations all asymptote to a power-law for a given T, I because of the constraint on the total information.  So ergodic space is whittled down simply because many configurations are so rare, it would be extremely unlikely ever to see them.  It is not easily possible to judge how much this process whittles down what is likely to be observed.  Looking at Fig \ref{fig:mdata}, the canonical solution of equation (\ref{eq:minifst}), all the other solutions with an area of 1 under the curve which therefore satisfies the requirement of a probability distribution, have been eliminated in the sense of being very unlikely.  It is certainly clear that what is likely to be observed, (i.e. the CoHSI solution) is a tiny subset of what might under extreme circumstances be observed.  Quantifying this observation is challenging but it is entirely consistent with the simple calculation earlier in this section.

\subsection{How much of protein sequence space is documented?}
The study of \cite{Dryden2008} discussed above dealt with the theoretical extent of protein space.  Given that we cannot realistically access that sector of protein sequence space that was explored by extinct organisms, we can address only what can be observed in modern life forms.  This discussion is normally couched in terms of species rather than proteins.  For example, \cite{Mora2011} estimated that 86\% of land species and 91\% of marine species remain to be described, so we have perhaps sampled some 10\% of the total species thought to be extant; we have concerns over these numbers that we will consider at the end of this section.  However a discussion based on species is not obviously the same as a discussion based on proteins because we would need to know how proteins are distributed by species.  This we can explore using CoHSI.

\subsubsection{Viewpoint using proteins}
A basic property of a CoHSI system is that the more components there are, the more emphatic is the CoHSI signal.  For smaller numbers the signal is visible but for large numbers it is completely dominant as we saw in the discussion leading from Fig. \ref{fig:tdata} to Figs \ref{fig:archaea_eps} - \ref{fig:drosi_eps}.  This is because when we choose proteins as components with tokens of amino acids to generate the CoHSI length distribution, then the numbers are indeed large, for example, there are around 80 million proteins with a median length of around 300 amino acids in TrEMBL 17-03.  CoHSI is then completely dominant with a demonstrably precise power-law in the tail of both the length and alphabet distributions \cite{HattonWarr2015}.

The point of this discussion is that we are trying to get to a theoretical relationship between species and proteins.  Following the success of using proteins as components and amino acids as tokens in revealing the CoHSI length distributions, \textit{suppose then we now use species as components and proteins as their tokens}.  We note that the definition of components and their token contents is only a matter of categorisation - providing it is consistent, CoHSI is still applicable, \cite{HattonWarr2017}.  We are free to do this provided we are consistent, as this is how we humans categorise and then measure the universe we live in.  As a result we expect species and proteins also to be related by a CoHSI distribution but with rather more noise present as there are self-evidently many fewer species in TrEMBL than there are proteins.

Before continuing with this argument, we should take a moment to consider the non-trivial problem of extracting species names from the TrEMBL releases and define what we mean by a species.  This will become important when we attempt below to associate TrEMBL protein data, using the species name, with a phylogenetic dataset.  Paraphrasing the Uniprot documentation (https://web.expasy.org/docs/userman.html, https://www.uniprot.org/docs/speclist), there are \textit{real} organism codes used in both Swissprot and TrEMBL and \textit{virtual} organism codes used only in TrEMBL. Real organism codes correspond to a specific and specified organism.  Virtual organism codes regroup organisms at a certain taxonomic level and generally correspond to a 'pool' of organisms, which may be as wide as a kingdom, as it is not possible in a reasonable timeframe to manually assign organism codes to all species represented in TrEMBL.  Instead Uniprot assigns a specific "official" name to an aggregated group of organisms such as Amphibia. Bioinformatically, Uniprot and TrEMBL releases identify an organism by a card image in the form:-

\begin{verbatim}
ID X_Y ...
\end{verbatim}

In the Swissprot release, X is a protein name and Y is a real organism code.  In TrEMBL, X is the accession number (https://web.expasy.org/docs/userman.html\#AC\_line) and Y the organism code (real or virtual).  The virtual codes all begin with the number 9.  There are some 23,204 organism records in TrEMBL 18-02 each identified by an "ID X\_Y" card image, and of these 285 are virtual; some of these virtual organisms contain many related species and in some cases correspondingly large aggregations of sequences.  The question is of course, how should we treat the virtual organism codes acting as they do, as a 'pool' of organisms which cross species boundaries?  Taxonomically, this would be very complex and many of them have very few proteins sequenced, which is why they are grouped in this way prior to the manual review winnowing process in Swissprot.  CoHSI however offers a unique simplification of this problem - it is agnostic with regard to the categories provided they are consistent. \textit{ We will simply define our species name as the value of the Y code on the "ID" card image, and extract the full scientific name from the "OS" or Organism Species card image (https://web.expasy.org/docs/userman.html\#OS\_line).}

This approach using our definition of species as components and proteins as tokens might seem to be in conflict with our previous analyses using proteins as components and amino acids as tokens. This is not the case, since we already know that the same discrete system can house multiple concurrent CoHSI systems, but with different unique alphabets of categorisation.  For example, we illustrated this by demonstrating the presence of two heterogeneous CoHSI systems co-existing in digital music depending on whether or not we include musical note duration as a separate category over pitch \cite{HattonWarr2017}.  We also demonstrated a homogeneous CoHSI system when we count words in texts co-existing with a heterogeneous CoHSI system when we count letters in words in the very same texts.  (A homogeneous CoHSI system is one in which each component contains only tokens of the same type unique to that component.  A heterogeneous CoHSI system is one in which each component contains multiple types of token.)

\textit{We therefore consider a heterogeneous CoHSI system in which components are \textbf{species} as defined above, and tokens are \textbf{proteins}.  In this system, we expect the distribution of numbers of proteins in species also to obey the CoHSI distribution (with different $\alpha, \beta$ of course).}  This we can test and we use the largest dataset, TrEMBL for this purpose.  Furthermore, to test the agnosticism of CoHSI with regard to the categories, we will do this by both including and excluding the virtual organism records.

\begin{figure}[ht!]
\centering
\includegraphics[width=0.5\textwidth]{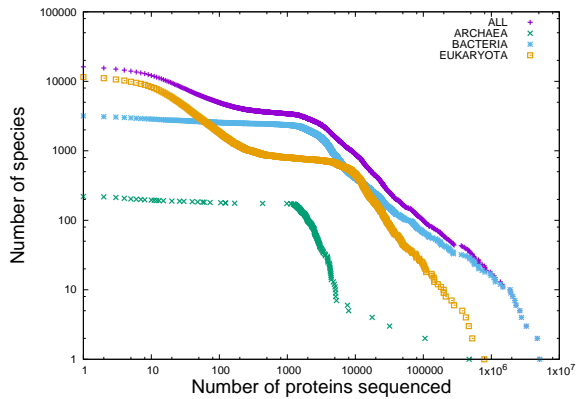}
\caption{\label{fig:speciesproteins}A ccdf of the distributions of number of proteins in species in the full TrEMBL 18-02 dataset for the domains of life, individual and in aggregate INCLUDING the virtual organisms.}
\end{figure}

\begin{figure}[ht!]
\centering
\includegraphics[width=0.5\textwidth]{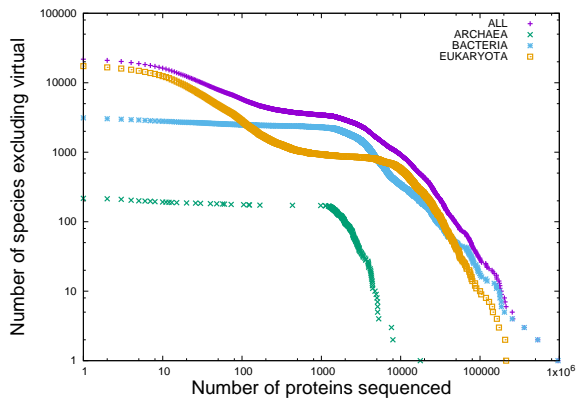}
\caption{\label{fig:speciesproteins_novirtual}A ccdf of the distributions of number of proteins in species in the full TrEMBL 18-02 dataset for the domains of life, individual and in aggregate EXCLUDING the virtual organisms.}
\end{figure}

Fig. \ref{fig:speciesproteins} is a ccdf of the distribution of numbers of proteins in species in the full release 18-02 of TrEMBL and shows the data broken down for the domains of life, i.e. archaea, bacteria and eukaryota \textit{including} the virtual organism records.  Fig. \ref{fig:speciesproteins_novirtual} is the same dataset but this time \textit{excluding} the virtual organism records.  We can note three things.

\begin{enumerate}
\item The tails of the distributions show the linearity expected from the predicted power-law, reflecting the insensitivity of CoHSI to the categorisation provided it is consistent; however, the tails of the distributions have different slopes.
\item When the virtual organisms are excluded, the slopes of the tails of the distributions are more consistent.
\item In the eukaryotic distributions there is a noticeable kink or plateau between around 10 and 10,000 sequenced proteins  which is not present in either the archaea or bacteria.  Given that this kink appears in \textit{both} Fig. \ref{fig:speciesproteins} and \ref{fig:speciesproteins_novirtual}, this is probable evidence of researcher bias in exploring the vastness of life.  We note that this bias does not appear in Fig. \ref{fig:swissprottrembl} which (because proteins are the components in this analysis) contains many more datapoints, thereby strengthening the CoHSI signal. We will now look at this probable researcher bias in a little more detail.
\end{enumerate}

First of all, if we took a minimum qualifying number of proteins even as low as 500 in order to measure how well researchers have covered a particular species (using our definition of species above of what we are actually measuring), then more than 98\% of the species appearing in the better curated SwissProt dataset would \textbf{not} qualify.  Even if the very much larger TrEMBL v 18-02 is used (larger by about a factor of 100x), almost 85\% of the species would not qualify either.

Another manifestation of this, at least currently, is that the number of species real or virtual, with sequenced proteins appearing in the full TrEMBL database is growing much more slowly than the number of proteins.  Table \ref{tab:tremblcontents} shows that in a space of some 20 months between two recent versions of TrEMBL, the total number of proteins increased by over 60\% but the number of species (real and virtual) increased by less than 4\%.  In other words current emphasis appears to be almost exclusively on making the proteomes of species which are already in TrEMBL more complete.  It is hard to predict how long this differential bias will persist.

\begin{table}[h!]
\centering
\caption{Numbers of species and proteins in versions of TrEMBL}
\begin{tabular}{@{\vrule height 10.5pt depth4pt  width0pt}cccc}
\hline
TrEMBL version & Species & Proteins & Average proteins per species \\
\hline
v. 15-07 & 22,030 & 49,401,998 & 2,242 \\
v. 17-03 & 22,824 & 80,204,459 & 3,514 \\
\hline
\end{tabular}
\label{tab:tremblcontents}
\end{table}

Finally, it is clear that an incomplete picture of proteome size for species in the TrEMBL database is presented in Fig. \ref{fig:speciesproteins_novirtual} (since the analysis was not restricted to organisms with whole genomes sequenced).  We know that when there were only 63 million proteins in TrEMBL from  \cite{doi:10.1093/nar/gkw978}, that the mean size of proteomes, i.e. mean number of sequenced proteins per species for viruses, archaea, bacteria and eukaryota was 42, 2358, 3200 and 15145 respectively suggesting that some flattening of the curve for species in the lower ranges of sequenced proteins might be expected. Fig. \ref{fig:speciesproteins_novirtual} provides substance for this but with some subtleties.  For example, we note that this figure appears to support our predictions about the CoHSI asymptotic state in that both archaea (1\% of all proteins in TrEMBL) and bacteria (48\% of all proteins in TrEMBL) appear to exhibit the least researcher bias as their curves are closest to the predicted CoHSI asymptotic ccdf shape \cite{HattonWarr2018a}.  In contrast the eukaryota (47\% of all proteins in TrEMBL) show a substantial dip between 10 and 10,000 proteins, representing primarily those eukaryotic species currently with a significant shortfall in the sequencing of their proteomes.  This shortfall observed in the eukaryotes is also manifest in a smaller but similar dip in the curve for the full TrEMBL dataset.  \textit{We can therefore make the (falsifiable) prediction that in the fullness of time the eukaryote dataset (and full TrEMBL dataset) will morph into the asymptotic CoHSI ccdf already observed in Fig \ref{fig:speciesproteins_novirtual} for the archaea and bacteria.}

So we can indeed observe a relationship, guided by CoHSI, between species and proteins  (and as a beneficial side-effect, we can observe the effects of researcher bias).  It is therefore reasonable to consider from this different perspective (i.e. arguments focused upon species) the central question of this section - the degree to which protein sequence space is documented.

\subsubsection{Viewpoint using species}
 To close this section, we reflect on the above estimates of how far we have to go in sequencing the proteins of life by considering species.  \cite{Mora2011} reported that 86\% of land species and 91\% of marine species remained to be described.  In terms of the actual numbers of species they estimated  "∼8.7 million (+/-1.3 million SE) eukaryotic species globally, of which ∼2.2 million (+/-0.18 million SE) are marine. In spite of 250 years of taxonomic classification and over 1.2 million species already catalogued in a central database, our results suggest that some 86\% of existing species on Earth and 91\% of species in the ocean still await description." It is worth noting that these authors give estimates of species numbers that are very much lower than those reported by others. Locey and Lennon (\cite{Locey201521291} applying ecological scaling laws estimated upwards of one trillion species on earth, with the majority  being bacteria. Larsen et al. (\cite{Larsen2017} estimated the total number of species to be 1-6 billion, with up to 90\% of these being bacterial species.  

Even if we accept the lowest of these estimates (\cite{Mora2011} ) of the total number of species, \textit{only 0.2\% have been sequenced, and these only partially,} and yet these already take up 18.5 Gigabytes when optimally compressed!  This suggests that to sequence all of the known lifeforms on earth will take some 18.5 x 500 Gigabytes $\sim 10$ Terabytes x $F$ where $F$ is whatever factor by which we decide the current TrEMBL database is undersampled for the species it already contains.  Given the rate of change in protein addition compared with species addition, a factor of 10 is probably very conservative but nevertheless leads to an estimate of perhaps 100 Terabytes in an optimally compressed form, which uncompressed, would amount to some 350 Terabytes. Even at this level, a tiny fraction of the potential space would have been explored by all the protein sequences on earth.

However, we can put even these huge estimates into context by noting that terrifying though this prospect may sound, these numbers are already dwarfed by the amount of data acquired by the Large Hadron Collider, which in July 2017 surpassed 200,000 Terabytes\footnote{https://home.cern/about/updates/2017/07/cern-data-centre-passes-200-petabyte-milestone, accessed 19-Oct-2018}.

\subsection{What has been the effect of mass extinctions?}
During the largest of the five commonly documented mass extinctions, the Permian-Triassic, some estimates suggest that 90\% of all species disappeared \cite{Benton2005}.  Even though gene sharing between species through for example, the mechanism of horizontal gene transfer, whereby proteins shared amongst species could survive if some but not all of those species disappeared, it seems unavoidable that this extinction would have reduced the number of proteins substantially, although it has been argued that the biomass associated with complex multicellular eukaryotes has remained approximately stable \cite{Franck2006}.  We would anticipate that after such an extinction carried out sufficiently rapidly, that the equilibrium relationship we describe between the total number of proteins and the longest might take some time to re-establish.  Of course we have no modern data on this but taxonomic and ecological recovery have been described in detail \cite{Sahney2008}.  We will return to this shortly when we look for extinction footprints in the protein data.

\subsection{Piecing these factors together}
Given that CoHSI exerts its constraints regardless of how proteins might be distributed amongst species, we could reasonably argue that the total number of proteins might increase by several orders of magnitude if we were able to sequence the entire population of species on the Earth, problems with continual extinction and speciation not withstanding.

Given also that the largest protein in the TrEMBL distribution is currently around 36,000 amino acids, using (\ref{eq:geometry}), this would suggest that the largest protein which might be discovered in the future may well be as large as $36000 (10^{1/3.13}) \sim 75,126$ amino acids.  Of course, this is a noisy estimate as explained earlier and in addition, there may well be biochemical and structural stability limits on such long proteins, although our discussion of protein folding did not throw up any obvious barriers. However, we can be a little more circumspect given the noise in the tail of longest proteins in the protein length distribution by extending our analysis using v. 17-03 of TrEMBL as a convenient baseline with regard to the point beyond which there are only 1,000 proteins, which roughly corresponds to the $99.999^{th}$ percentile.  As can be seen by studying Fig. \ref{fig:tremblversions}, a value of 1,000 amino acids in length is comfortably within the self-similar linear tail in both distributions.   For TrEMBL release 17-03, the length corresponding to the $99.999^{th}$ percentile is 10,787 amino acids, which we will notate as $l_{1000}$.  In other words this is the length of the $1000^{th}$ longest protein.  We therefore extend this by using (\ref{eq:geometry}) with various very conservative scenarios as shown in Table \ref{tab:predictedbins} for the same value of $\beta' = 4.13$ used earlier.  The first row simply reproduces the observed results for version 17-03 of TrEMBL.
\begin{table}[h!]
\centering
\caption{Predicted bins of long proteins in the complete proteome, from a baseline of version 17-03 of TrEMBL.}
\begin{tabular}{@{\vrule height 10.5pt depth4pt  width0pt}cccc}
\hline
Factor increase on 17-03 & $l_{1000}$ & $l_{500}$ & $l_{100}$ \\
\hline
x1 & 10,787 & 13,580 & 21,538 \\
\hline
x5 & 18,039 & 22,710 & 36,018 \\
x10 & 22,511 & 28,339 & 44,946 \\
x15 & 25,624 & 32,259 & 51,163 \\
\hline
\end{tabular}
\label{tab:predictedbins}
\end{table}

\textit{Taking a guess of a x10 factor increase in total number of proteins over version 17-03 in the next few years, it seems likely that there will be a significant number of proteins (around 100) which are longer than about 45,000 amino acids in species yet to be discovered and/or sequenced.  We discuss later the domain of life in which this is most likely to occur.}

\section{The longest proteins and their evolutionary significance}
We have demonstrated both by theory and experiment in preceding sections that it is overwhelmingly likely that the longest protein in a collection of proteins is determined \textit{solely} by a) the total number of proteins in that collection and b) the power-law slope of the pdf of the length distribution, which we will call $\beta'$.  We have also shown that the value of $\beta'$ tends to be highly conserved, accounting for the self-similar nature of Figures \ref{fig:tremblversions} and \ref{fig:swissprottrembl}.  Given that the total number of proteins must by definition have grown from zero when life first appeared, although perhaps not monotonically as speciations and extinctions have occurred, \textit{we would expect our theory to show that the longest proteins will tend to occur in recent times, and might perhaps provide some evidence of the mass extinctions.}

\subsection{CoHSI and the recent emergence of long proteins}
In order to test this hypothesis, we merged protein data \textit{using only real organism codes} in version 18-02 of the TrEMBL dataset with the evolutionary data on the time of species emergence in the \textit{Time Tree of Life}\footnote{http://www.timetree.org/, accessed 07-Oct-2018}.  The Time Tree of Life is assembled as a phylogenetic tree and the TrEMBL dataset contains only those species for which some subset (potentially a large proportion) of their proteome has been sequenced.  Clearly the Time Tree of Life will contain species now extinct and which have no such proteome whereas the TrEMBL dataset can contain the proteome of ancient species but only if they have been extant in the last few thousand years and for which DNA is still accessible, (for example that of the Mammoth).  Merging the two datasets using the species name extracted from the "OS" header lines in the TrEMBL .dat format files for real organism codes, and the species name extracted from the phylogenetic data, gives a potential population of 9,469 species which appear in both the Time Tree of Life (file TimetreeOfLife2015.nwk, as of 28-Sep-2018) and also in version 18-02 of the TrEMBL dataset.

Before proceeding, we note that predicting the role of the longest protein depends of course on reliable estimates of that longest protein given that many species proteomes are incomplete.  There are in fact complete proteomes available in Uniprot, specifically documented as \textit{reference proteomes} (https://uniprot.org/proteomes), including for example  \textit{Homo sapiens}; however there are not very many of these so we have a typical statistical trade-off between completeness and size of dataset.  We therefore approach this in two ways.

\subsubsection{Estimating the longest proteins of a species}
\begin{figure*}[t!]
    \captionsetup[subfigure]{labelformat=empty}
    \centering
    \begin{subfigure}[t]{0.5\textwidth}
        \centering
        \caption{A}
        \includegraphics[width=6cm]{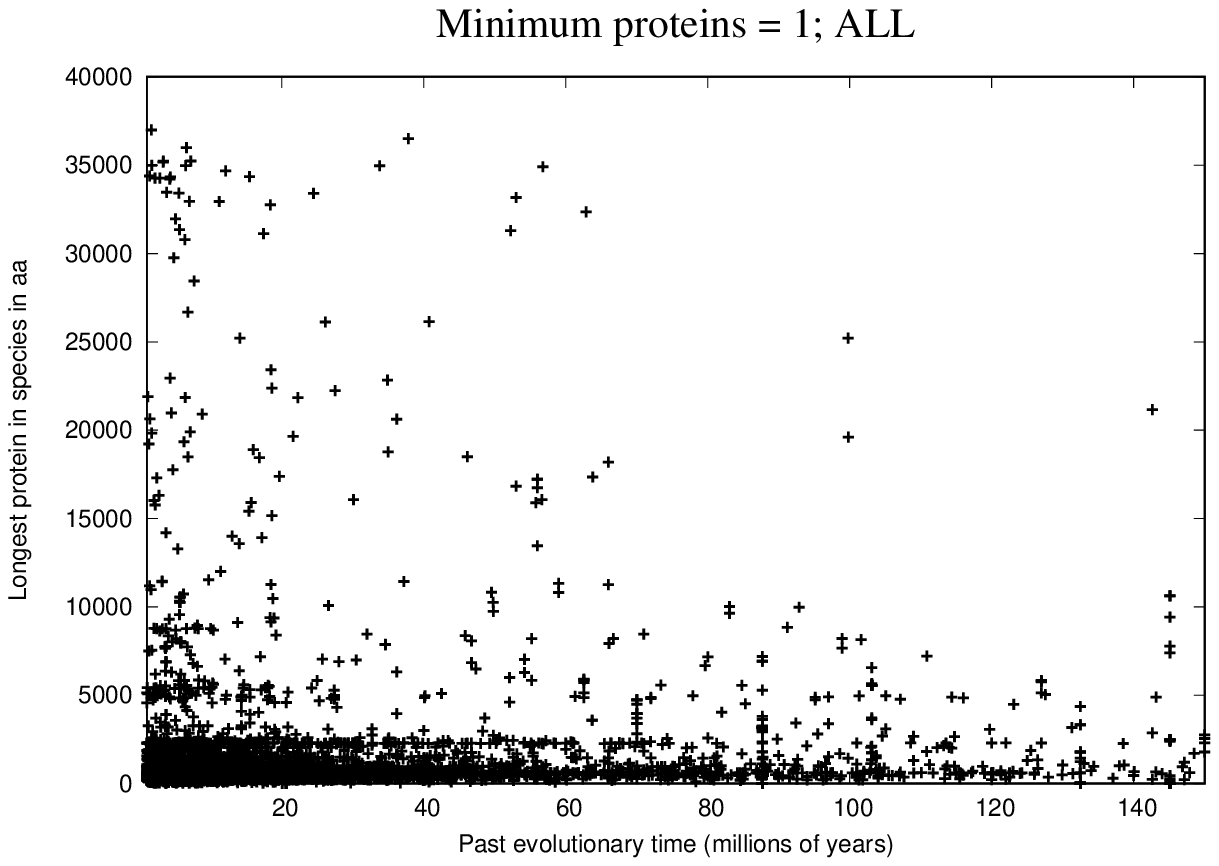}
        \label{fig:time_maxprot_1prot_150_ALL}
    \end{subfigure}%
    ~
    \begin{subfigure}[t]{0.5\textwidth}
        \centering
        \caption{B}
        \includegraphics[width=6cm]{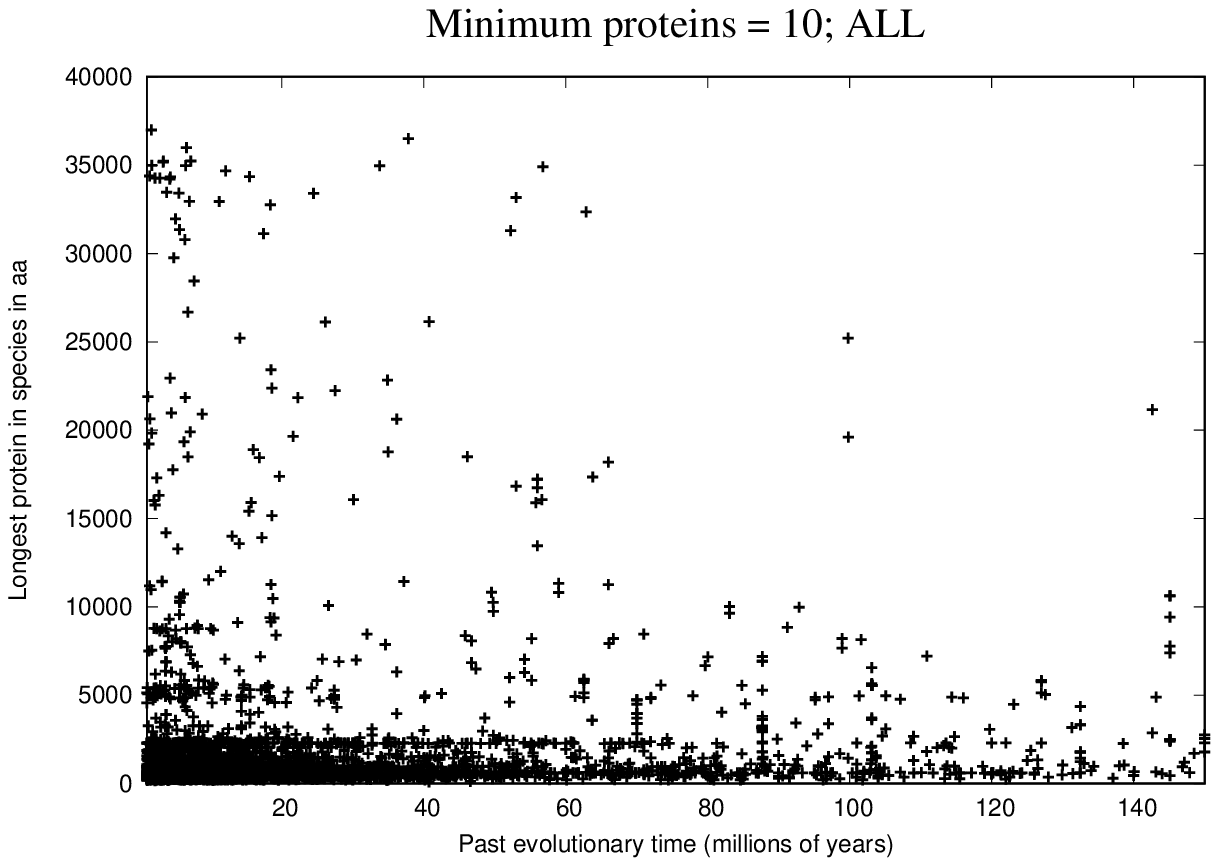}
        \label{fig:time_maxprot_10prot_150_ALL}
    \end{subfigure}%

    \begin{subfigure}[t]{0.5\textwidth}
        \centering
        \caption{C}
        \includegraphics[width=6cm]{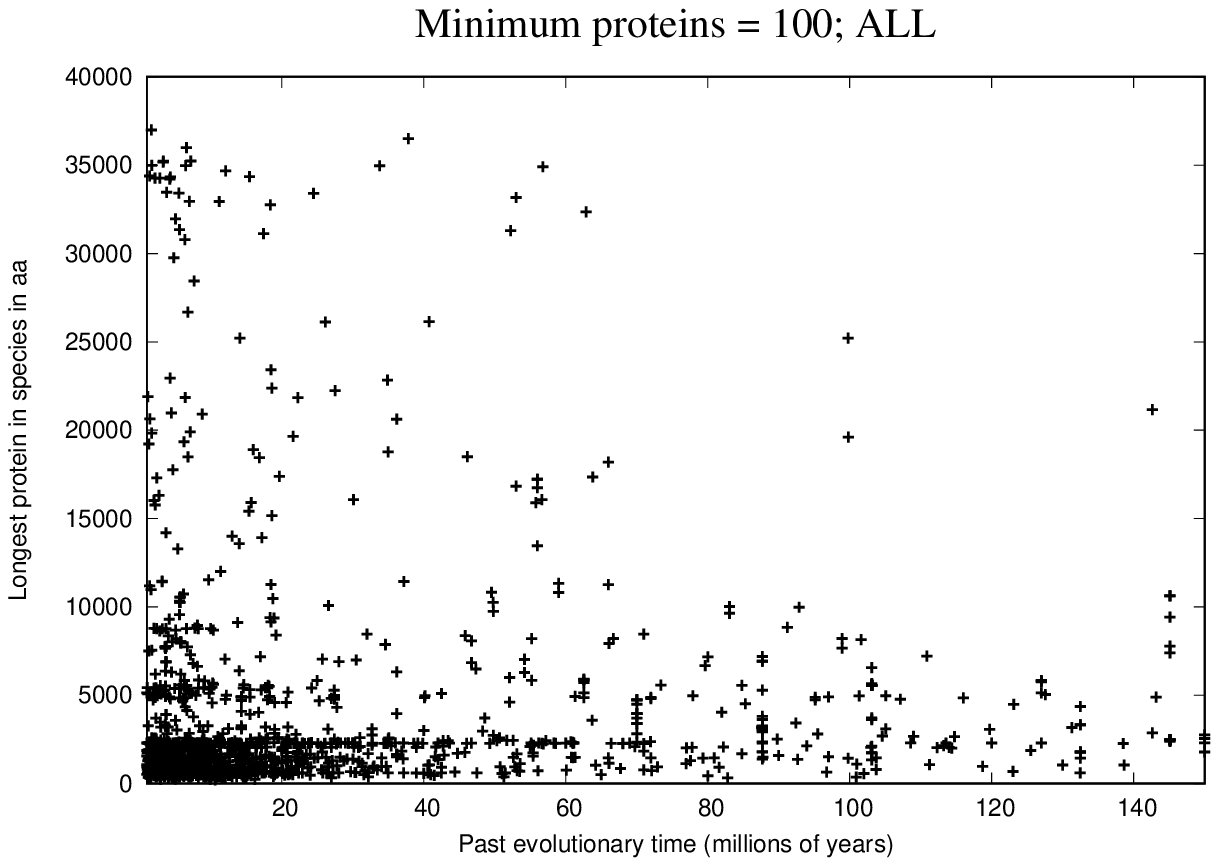}
        \label{fig:time_maxprot_100prot_150_ALL}
    \end{subfigure}%
    ~
    \begin{subfigure}[t]{0.5\textwidth}
        \centering
        \caption{D}
        \includegraphics[width=6cm]{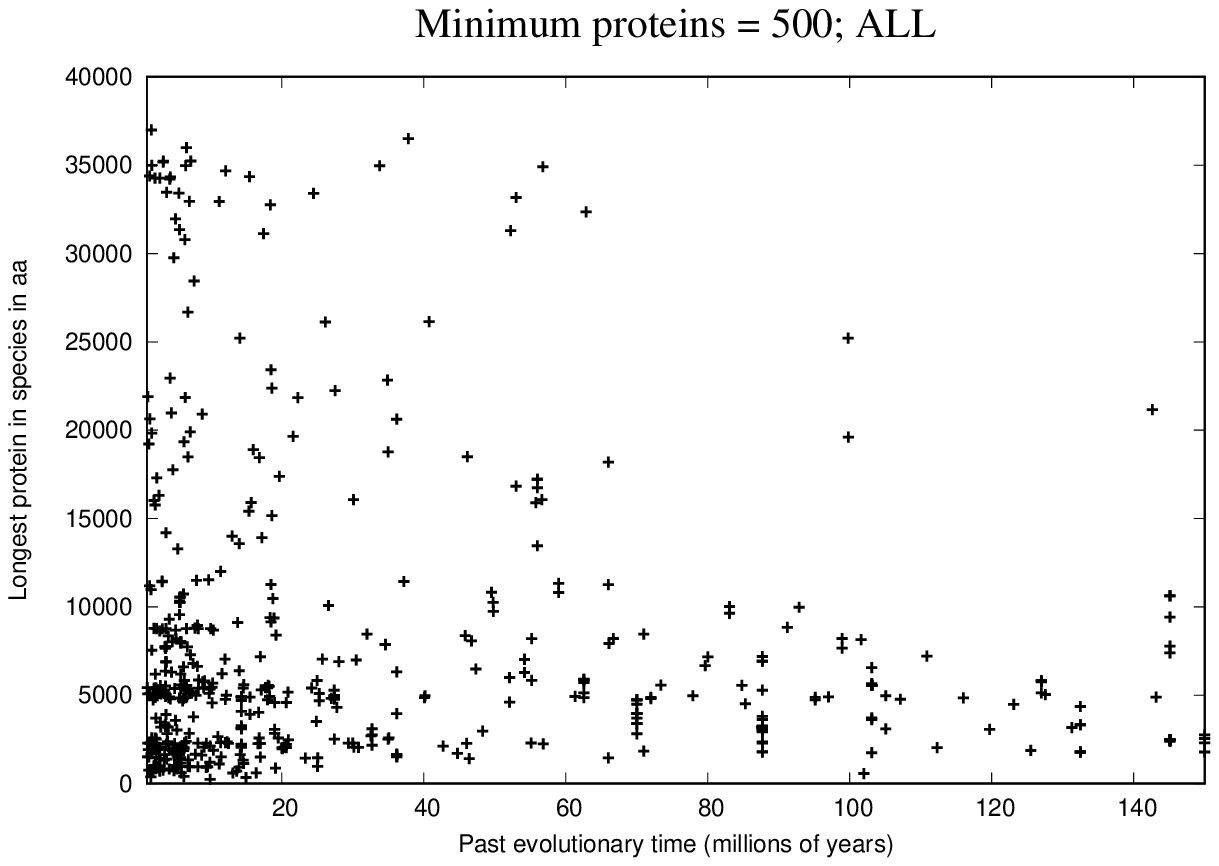}
        \label{fig:time_maxprot_500prot_150_ALL}
    \end{subfigure}%

    \begin{subfigure}[t]{0.5\textwidth}
        \centering
        \caption{E}
        \includegraphics[width=6cm]{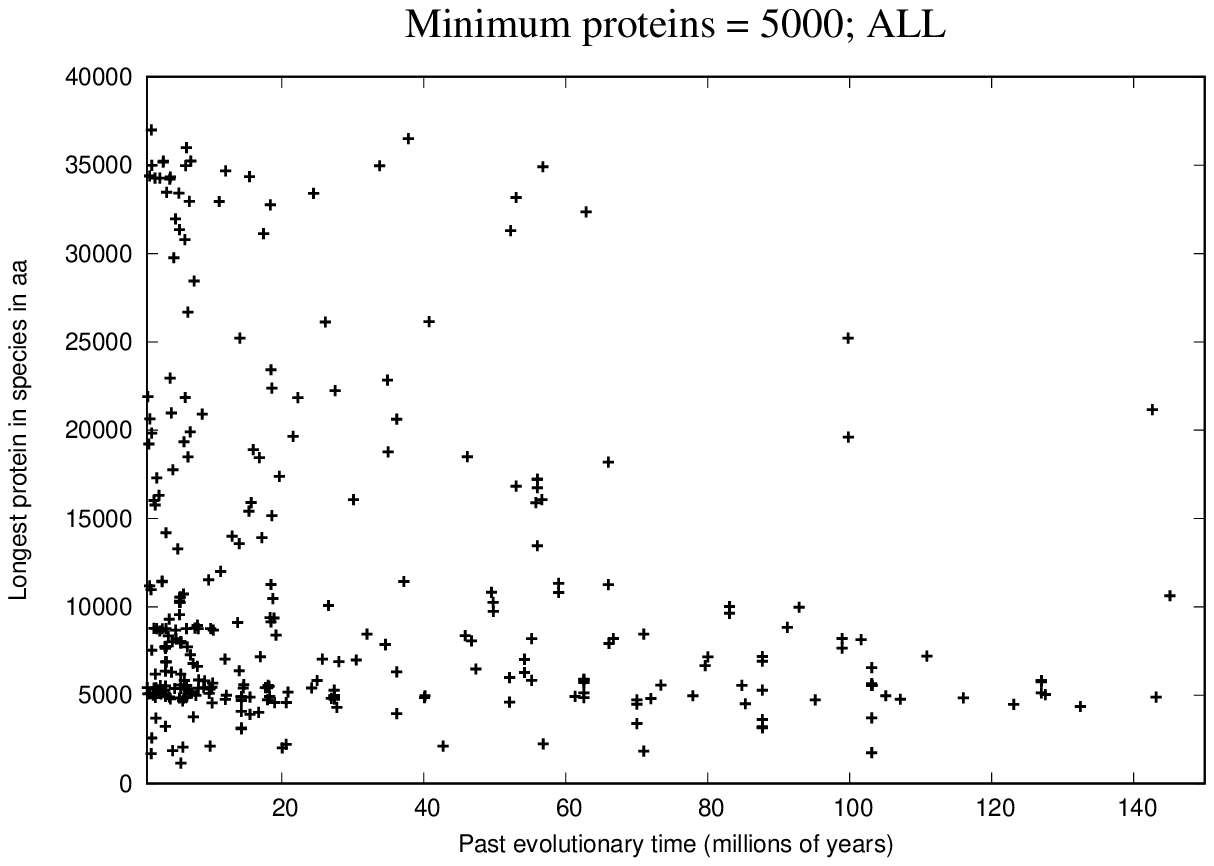}
        \label{fig:time_maxprot_5000prot_150_ALL}
    \end{subfigure}%
    ~
    \begin{subfigure}[t]{0.5\textwidth}
        \centering
        \caption{F}
        \includegraphics[width=6cm]{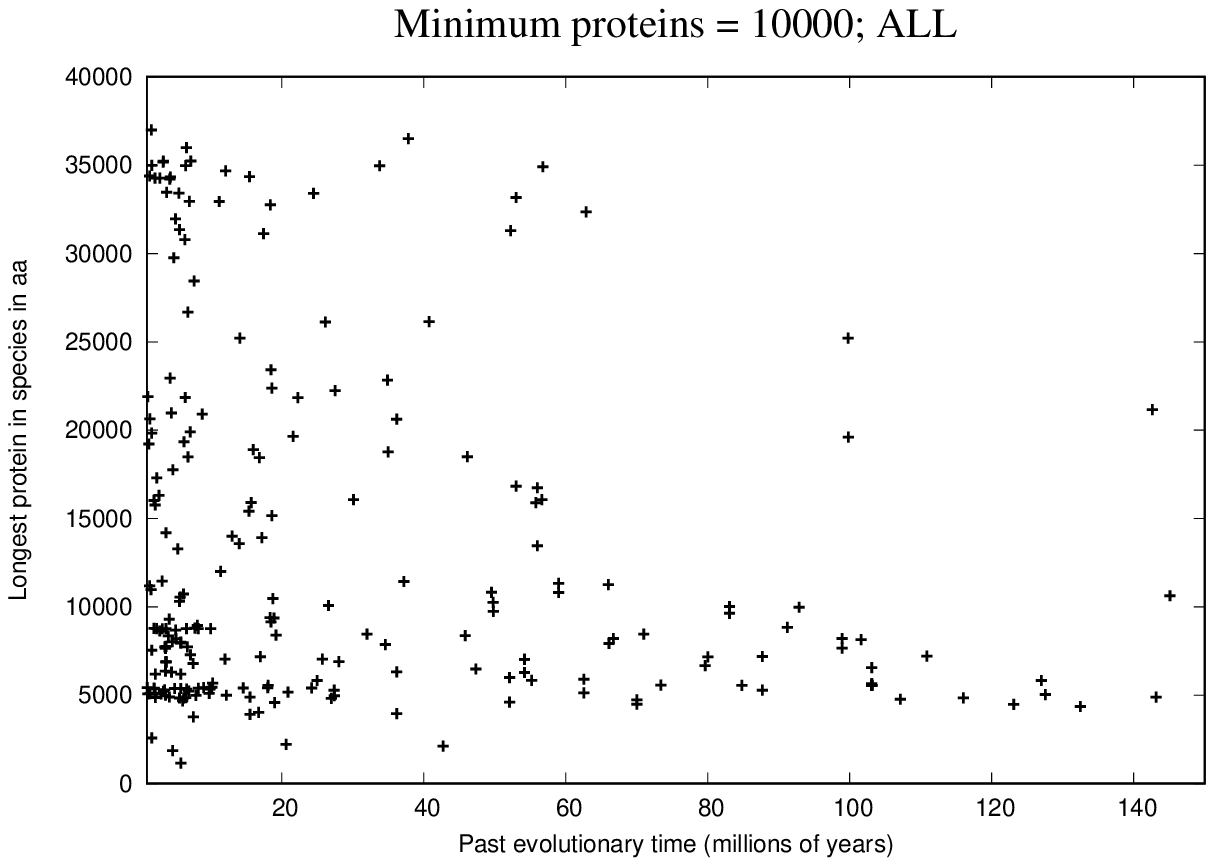}
        \label{fig:time_maxprot_10000prot_150_ALL}
    \end{subfigure}%
    \caption{Different qualifying proteome sizes, all illustrating the appearance of the longest protein against evolutionary time in the last 150 Mya: A) 1 protein; B) 10 proteins; C) 100 proteins; D) 500 proteins; E) 5,000 proteins; and F) 10,000 proteins.}
\end{figure*}

In order to get a reasonable estimate of the size of the longest proteins in a species, we should specify some minimum qualifying number of proteins in the species' proteome.  The appropriate size of the sample that should be available from a species genome is a difficult question to resolve statistically as we are confronted (amongst other challenges) by potential unknown researcher bias in acquiring the data. Thus we took an empirical approach, specifying a variety of minimum qualifying numbers (from 1 - 10,000) of proteins for a species to be included in the analysis.  Data for each qualifying species were extracted from  TrEMBL version 18-02,  and plotted against the time of evolutionary emergence.   Figures \ref{fig:time_maxprot_1prot_150_ALL} - \ref{fig:time_maxprot_10000prot_150_ALL} illustrate the observed relationships between the maximum length of protein in each qualifying species when the qualifying level for each species was set at 1; 10; 100; 500; 5,000; or 10,000 proteins, merged with the evolutionary time data from the Time Tree of Life.

\begin{figure*}[t!]
    \captionsetup[subfigure]{labelformat=empty}
    \centering
    \begin{subfigure}[t]{0.5\textwidth}
        \centering
        \caption{A}
        \includegraphics[width=6cm]{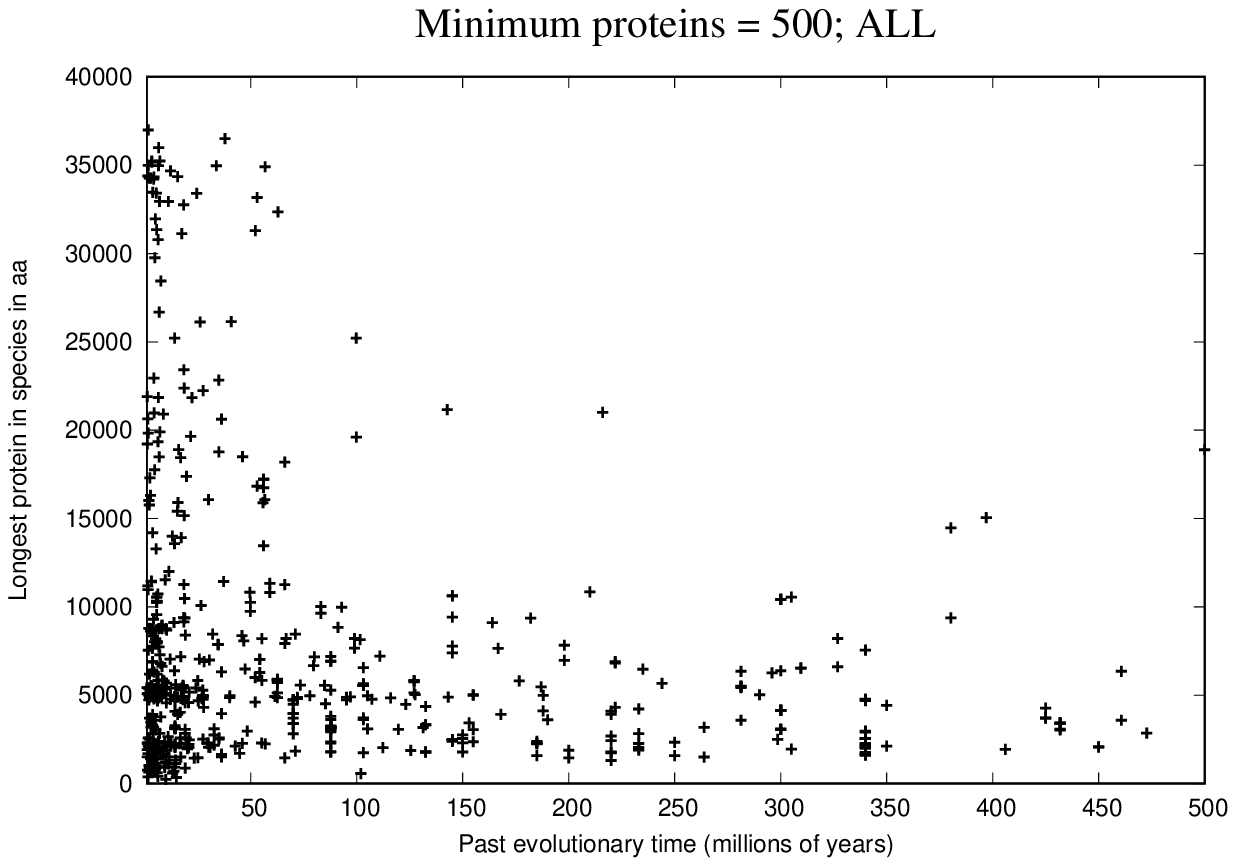}
        \label{fig:time_maxprot_500prot_500_ALL}
    \end{subfigure}%
    ~
    \begin{subfigure}[t]{0.5\textwidth}
        \centering
        \caption{B}
        \includegraphics[width=6cm]{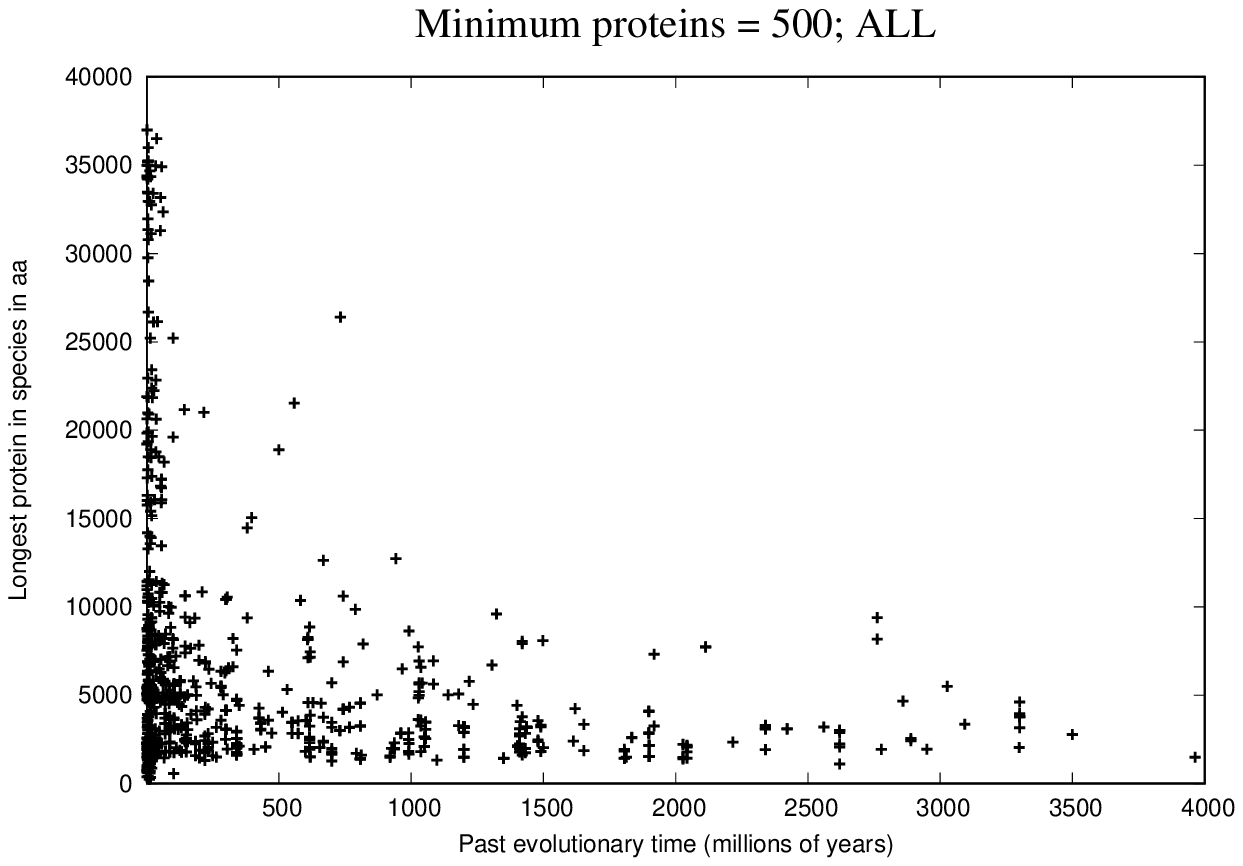}
        \label{fig:time_maxprot_500prot_4000_ALL}
    \end{subfigure}%
    \caption{Illustrating the appearance of the longest protein against evolutionary time: A) 500 Mya; and B) 4,000 Mya.}
\end{figure*}

Studying these results, horizontal linearities at low values on the Y-axis are obvious for minimum qualifying numbers of 1, 10, and 100 proteins; these suspected artifacts disappear progressively as the number of qualifying proteins/species increases, and we interpret these linearities as reflecting the statistical likelihood that a random small sample of proteins from a species will contain predominantly the relatively shorter and more numerous proteins (see Fig \ref{fig:tdata}).  Apart from the linearities, the data for all numbers of qualifying proteins look broadly similar, with a clear trend for the longest proteins in a species to correlate inversely with the time since its emergence. Thus we chose 500 proteins as our minimum size to qualify a species' proteome for further analysis in the context of time of evolutionary emergence. However, we stress that until we consider reference proteomes, this is an arbitrary decision. Considering only species with at least 500 proteins sequenced leaves 954 qualifying species.  Figures \ref{fig:time_maxprot_500prot_500_ALL} and \ref{fig:time_maxprot_500prot_4000_ALL} show these data for all qualifying species in TrEMBL version 18-02 whose times of evolutionary emergence are estimated up to 500 Mya or up to 4,000 Mya.

With a view to attempting to falsify the CoHSI predictions in the spirit of Popperian analysis, we note from Figures \ref{fig:time_maxprot_500prot_500_ALL} and \ref{fig:time_maxprot_500prot_4000_ALL}, that two things become clear.  First of all, the growth in the longest protein as evolutionary time winds forwards to the present day is obvious,  exactly as predicted by CoHSI.  Second, there are intriguing minima in the signal at around  80 Mya, 120 Mya, 250 Mya, 250 Mya, 430 Mya, 1700 Mya and 2500 Mya, and an obvious interpretation is that these reflect extinction events.  Before discussing these further however, we will gain further insight using the reference proteomes of Uniprot.

\subsubsection{Using reference proteomes}
\begin{figure*}[t!]
    \captionsetup[subfigure]{labelformat=empty}
    \centering
    \begin{subfigure}[t]{0.5\textwidth}
        \centering
        \caption{A}
        \includegraphics[width=6cm]{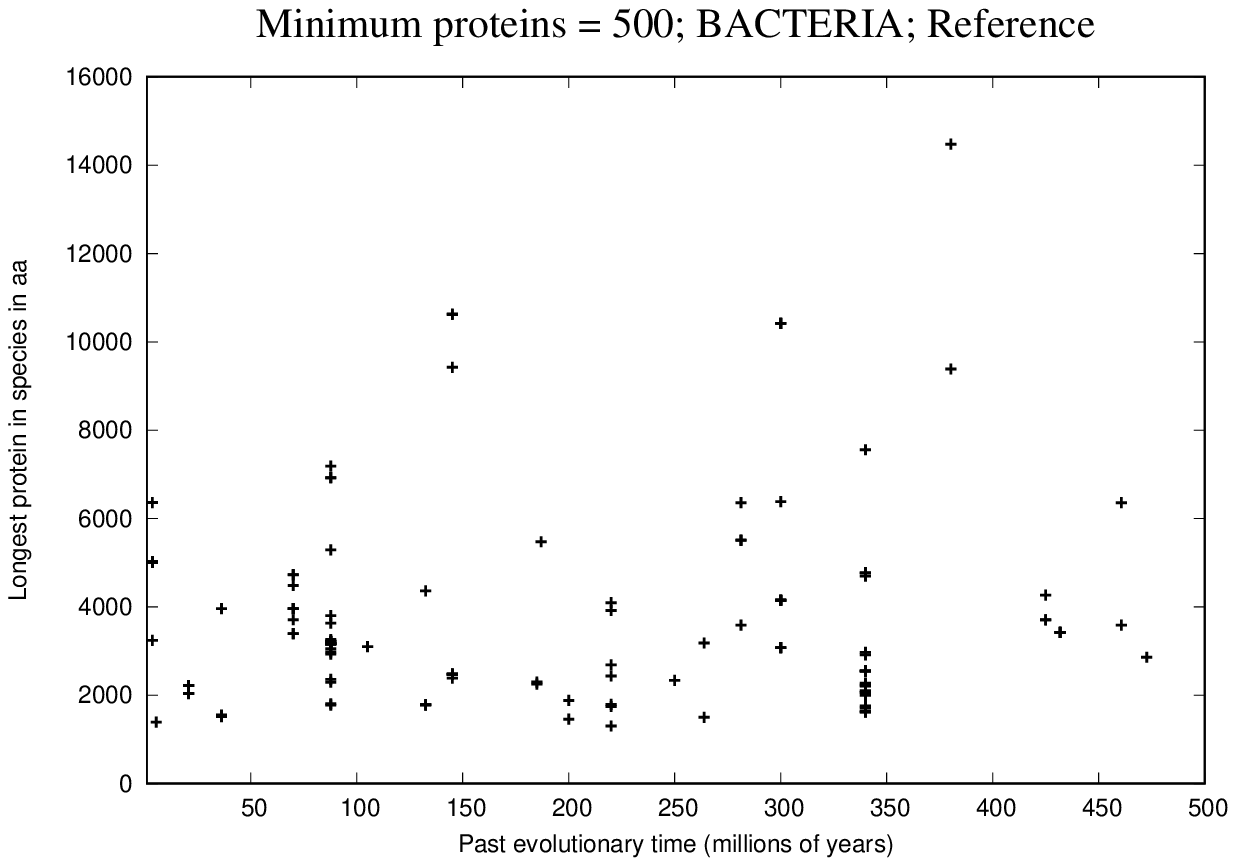}
        \label{fig:time_maxprot_500prot_500_BACTERIA}
    \end{subfigure}%
    ~
    \begin{subfigure}[t]{0.5\textwidth}
        \centering
        \caption{B}
        \includegraphics[width=6cm]{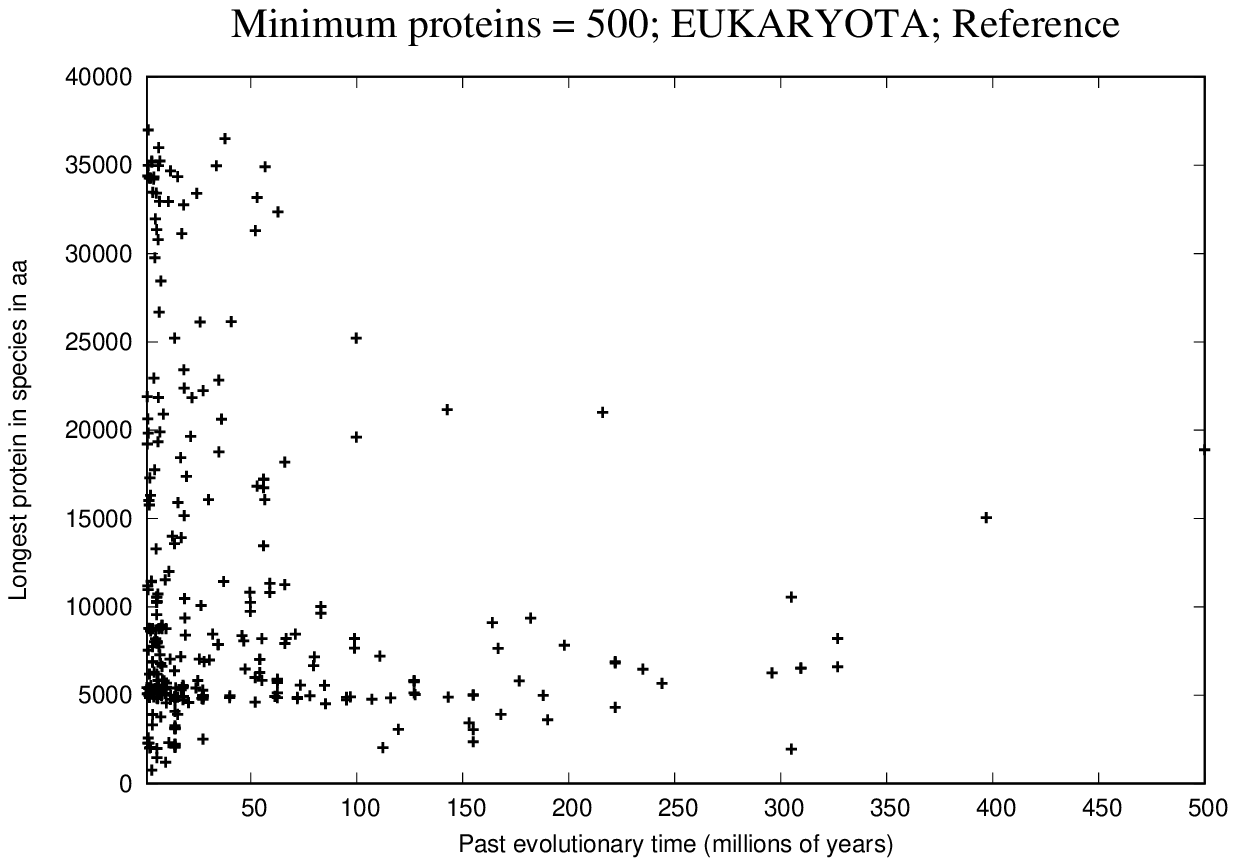}
        \label{fig:time_maxprot_500prot_500_EUKARYOTA}
    \end{subfigure}%
    \caption{Illustrating the appearance of the longest protein against evolutionary time in the last 500 Mya in: A) Bacteria; and B) Eukaryota by merging only the reference proteomes of Uniprot with the phylogenetic data.}
\end{figure*}

Certainly at first glance, the data of Figures \ref{fig:time_maxprot_500prot_500_ALL} and \ref{fig:time_maxprot_500prot_4000_ALL} support the predictions of CoHSI made earlier, i.e. that as the total number of proteins increases so does the size of the longest proteins.  Could there be other explanations for this ?  Looking deeper, it is certainly possible that the results seen in Fig \ref{fig:time_maxprot_1prot_150_ALL} - \ref{fig:time_maxprot_10000prot_150_ALL} may simply reflect researcher bias, but we can test and resolve this by looking at the same evolutionary timeline but this time for different domains of life and also by using only the \textit{reference proteomes} of Uniprot thereby eliminating concerns about estimating the longest protein in a species proteome from an incomplete subset.  Reference proteomes are complete sequenced proteomes for various chosen species, (around 10,000 currently for bacteria and eukaryota of which 705 also have estimated times of evolutionary emergence).  The results of merging the data on longest proteins with times of evolutionary emergence for these species with reference proteomes are shown in Figures \ref{fig:time_maxprot_500prot_500_BACTERIA} and \ref{fig:time_maxprot_500prot_500_EUKARYOTA} for the eukaryota and bacteria.

Figures \ref{fig:time_maxprot_500prot_500_BACTERIA} and \ref{fig:time_maxprot_500prot_500_EUKARYOTA} present intriguing results and are consistent with our estimation method which led to Figs \ref{fig:time_maxprot_500prot_500_ALL} and \ref{fig:time_maxprot_500prot_4000_ALL}.  It is clear from Figure \ref{fig:time_maxprot_500prot_500_BACTERIA} that the growth in longest proteins in recent evolutionary time is an exclusive property of the eukaryota, and is \textit{not} shared by the bacteria.  This seriously undermines the possibility of a consistent researcher-induced bias due to incompleteness.

However, it might also be said that since CoHSI predicts that the longest protein is simply a function of the total number of proteins, the more numerous bacteria (which represent a greater share of protein sequences, over 60\% in TrEMBL version 18-02) should therefore show a similar and perhaps even bigger growth in longest proteins as do the eukaryotes.  However, CoHSI also shows that the longest protein is a function of not only the total number of proteins \textit{but also the decay rate of its power-law, $\beta'$}.  Thus, within the overall CoHSI distribution, we predict that the subset of bacterial proteins must have a value of $\beta'$ distinct from that of the eukaryotic protein distribution. How can we test this?  We have discussed in previous work that the decay rate $\beta'$ is a complex function of the distribution of unique alphabets of amino acids found in each collection \cite{HattonWarr2017,HattonWarr2018a} and thus we could falsify CoHSI if the two domains of life, bacteria and eukaryota are \textit{indistinguishable} in their average unique amino acid alphabets.  This in turn would imply the same $\beta'$ in both domains and therefore that we should have seen an even more emphatic growth in long proteins in bacteria as compared to the eukaryota, given the greater abundance of proteins in bacteria.  Since this is clearly not the case, we test if the eukaryota and bacteria do indeed have the same average unique amino acid alphabet.

To do this we take all bacterial and eukaryotic proteins greater than 1,000 amino acids in length, which places them squarely in the power-law tail, and compare their average unique amino acid alphabets using the better annotated SwissProt version 18-02 which has more detailed information on post-translationally modified (PTM) amino acids, (which of course are instrumental in increasing the size of the unique alphabet \cite{HattonWarr2015} since there is an absolute maximum of 22 amino acids directly encoded by the genome).  Both the Kolmogorov-Smirnov test ($D = 0.52633, p < 2.2 \times 10^{-16}$) and a Welch two sample t-test ($t = -85.813, df = 15838, p < 2.2 \times 10^{-16}$) emphatically reject the null hypothesis that the sample means are the same. In fact, the unique amino acid alphabet of the eukaryota exceeds that of the bacteria by around 5.8\%, (21.18 v. 20.02), which we interpret as showing the greater influence of PTM in the eukaryota.

We note in passing that, in contrast to the information theoretic explanation, based on unique alphabet size, that we have presented (and tested experimentally) above, that a number of essentially biological arguments can also be adduced to explain the failure of the longest bacterial proteins to show an increase in size over evolutionary time. These arguments include, but are not limited to, consequences of the relatively small size and the highly crowded cytoplasm of typical bacteria \cite{McGuffee2010,Yu2016}, which impact properties including diffusivity, the frequency and strength of multiple weak interactions with neighboring macromolecules, and necessarily longer times for protein translation. However, since such arguments about the physicochemical properties of longer prokaryotic proteins are difficult to falsify in the Popperian sense, we feel it appropriate to reserve judgment on their possible relevance to the observations we have reported here.

We thus conclude this section by stating 1) that the CoHSI prediction of growth in the longest proteins in recent evolutionary time is entirely consistent with what we observe in the sequence databases. However, 2) our results suggest that the emergence of longer proteins is occurring predominantly in the most recently emerging eukaryotes; the converse of this is that once a species has emerged it is not likely to be the source of the emerging novel longest proteins. The concept that the evolutionary emergence of a taxon is associated with a lack of change at the molecular level is not borne out for mutation rates. Indeed, the concept that mutations accumulate at measurable rates (regardless of time of emergence) is the basis of the molecular clocks by which phylogenetic trees can be calibrated (\cite{Ho2014}).  \textit{This suggests that the emergence of novel very long proteins as predicted by CoHSI (and experimentally observed) operates in evolution independently of the well-understood processes of genetic mutation.}

\subsection{CoHSI and the extinctions}
Since our estimation method was essentially verified by using reference proteomes, we return to the discussion of Figures \ref{fig:time_maxprot_500prot_500_ALL} and \ref{fig:time_maxprot_500prot_4000_ALL}.  Minima or even gaps in the plot of maximum length of proteins against evolutionary time are clearly visible.  It is reasonable to question whether or not these correspond to known extinction events. An obvious example in Figure \ref{fig:time_maxprot_500prot_4000_ALL} is the dip at 2,500 Mya that is coincident with the Great Oxygenation Event (https://en.wikipedia.org/wiki/Great\_Oxygenation\_Event, accessed 08-Oct-2018).  A less obvious example is seen at around 1,700 Mya and this would be a candidate for a currently undocumented extinction event.

\begin{figure}[ht!]
\centering
\includegraphics[width=1.0\textwidth]{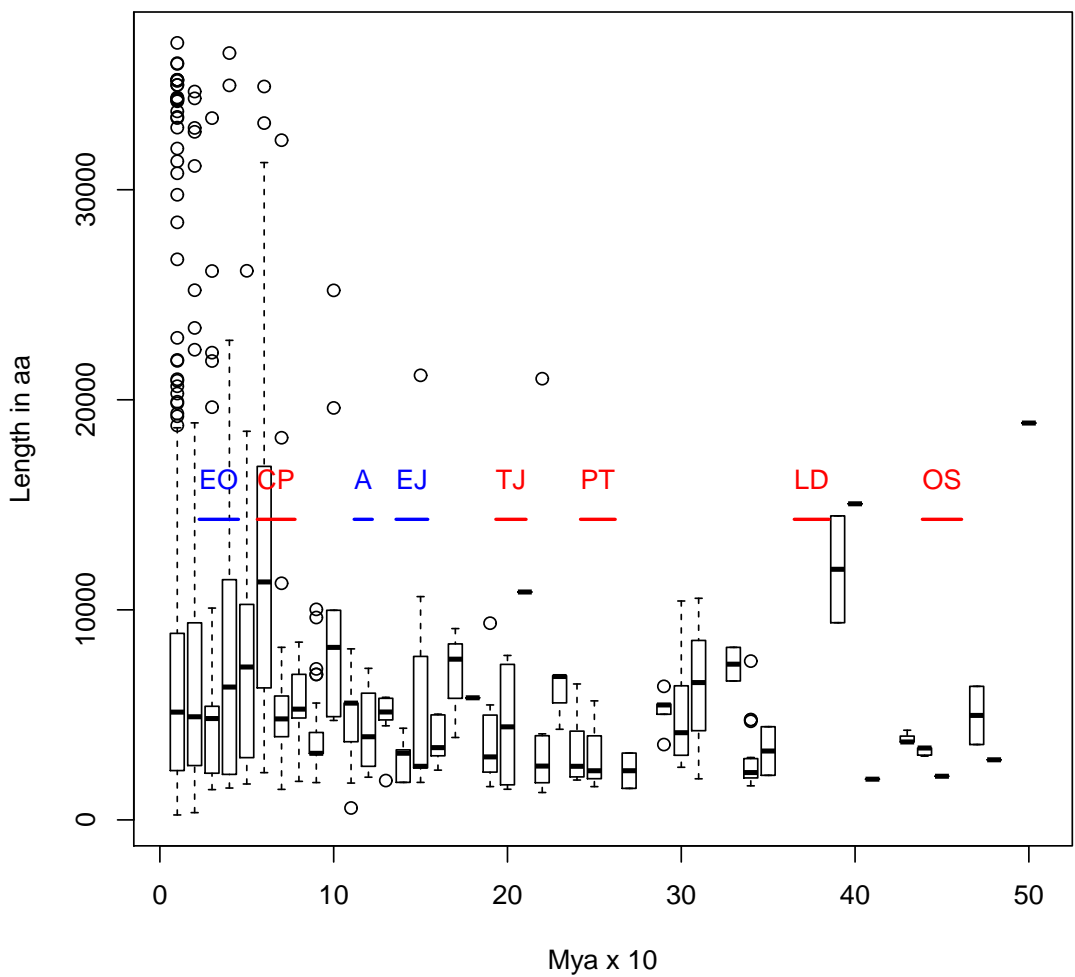}
\caption{\label{fig:boxplot}Illustrates the boxplots of the maximum protein lengths for all species in 10 Mya bins.  This is annotated with major extinctions in red and minor ones in blue.  Boxplots show the interquartile range with the median marked as a line.  The whiskers where shown indicate values within $1.5 \times$ the inter-quartile range and values outside this, known as outliers are shown as circles.}
\end{figure}

However to see the more recent extinction events of the last 500 Mya, we can consider the analyses shown in Figure \ref{fig:boxplot} where box and whisker plots of maximum protein lengths are given for all species within 10 Mya bins.  This plot is annotated with the documented major and some minor extinction events, as EO = Eocene - Oligocene, CP = Cretaceous - Paleogene, A = Aptian, EJ = End-Jurassic, TJ = Triassic-Jurassic, PT = Permian-Triassic, LD = Late-Devonian and OS = Ordovician-Silurian.  The red annotated ones belong to the normally described five major extinctions.  Each of the extinctions is shown with a line underneath showing the documented date of the extinction.  Although the indicated extinction dates are only approximate, there is an intriguing correspondence between these extinction events and a significant fall in the average longest protein across species contemporaneous with the event.  Whilst qualitative, the data are certainly not inconsistent with the CoHSI longest protein hypothesis, bearing in mind that the relationship between species number and protein numbers is not trivial as we described earlier in section 5.3.2.

\section{Conclusions}
The most significant point emerging from this paper is that the asymptotic power-law nature of the CoHSI length distribution means that the longest protein of any collection of proteins (e.g. a single species' proteome, a pan-proteome embracing multiple species, or the full collection of proteins in a database such as TrEMBL) is intimately and indeed simply related to the total number of proteins in that collection, through the parameter $\beta'$.  In other words as the collection grows in the number of proteins, the longest protein to be found in that collection grows in an entirely predictable way.  This process borrows nothing from evolution.  It is an information theoretic property of any discrete system which obeys the CoHSI distribution and we know through various preceding papers \cite{HattonWarr2017,HattonWarr2018a} that the known full collection of proteins in the TrEMBL and SwissProt databases very accurately obeys this length distribution.  We explored the consequences of this in several ways.

\begin{itemize}
\item Using this relationship and version 17-03 of the full TrEMBL distribution as a baseline, we have predicted under various scenarios how many proteins longer than some target figure are likely to appear in the months and years to come.  For example, if the total number of proteins grows by a factor of 10 within 5-10 years (fairly likely given that sequences have probably been obtained from fewer than 0.2\% of the current lifeforms on earth, and the known current rate of growth in TrEMBL), then we predict that approximately 100 proteins will be found longer than about 45,000 amino acids, either in existing species whose proteomes have not yet been fully sequenced, or in species yet to be sequenced.  We also pointed out that current researcher bias seems to be most obvious with the eukaryota because, in the analysis using species as components and proteins as tokens, they depart from the predicted CoHSI asymptotic state (Fig \ref{fig:speciesproteins_novirtual})  by a larger amount than the archaea or bacteria. Researcher bias also currently appears to favor completing the full proteome sequencing of existing species rather than investigating new species.
\item We then explored the degree to which sequence space has been populated by evolution on earth based on previous authors' studies, and suggest that only a tiny percentage of possible space has indeed yet been explored, although admittedly we have no means of assessing the contribution of extinct species.  Of course we can make no comment on how much more of this space has been populated by any extraterrestrial life with amino acid-based proteins, although we were able to give some insights as to how CoHSI and its backbone methodology from Statistical Mechanics explores the all-encompassing ergodic space.
\item Finally, we merged phylogenetic data and TrEMBL data to explore an intriguing prediction by CoHSI that the longest proteins should have appeared recently in evolutionary history, and that there should be traces of the earth's extinction events in these merged data.  These two predictions were well supported by the merged data.
\end{itemize}

In summary, we have shown that another property of CoHSI systems, the inevitability of very long proteins, manifests itself in a number of interesting ways in real data.  Each of these analyses further supports the fundamental role we believe CoHSI plays in setting bounds on the evolution of life.

\bibliographystyle{alpha}
\bibliography{bibliography}

\end{document}